\documentclass[lettersize,journal]{IEEEtran}
\usepackage[utf8]{inputenc}
\usepackage{url}

\usepackage{caption}
\usepackage{subcaption}
\usepackage{graphicx}
\usepackage{color}
\usepackage{amsmath,amsfonts}
\usepackage{algorithmic}
\usepackage{algorithm}
\usepackage{array}
\usepackage{textcomp}
\usepackage{stfloats}
\usepackage[english]{babel}
\usepackage{amssymb}
\usepackage[table]{xcolor}
\usepackage{makecell}
\usepackage{enumitem}
\usepackage{cite}
\begin{document}

\title{Vetaverse: A Survey on the Intersection of Metaverse, Vehicles, and Transportation Systems}


\author{Pengyuan Zhou, Jinjing Zhu, Yiting Wang, Yunfan Lu, Zixiang Wei, Haolin Shi, Yuchen Ding, Yu Gao, Qinglong Huang, Yan Shi, Ahmad Alhilal, Lik-Hang Lee, Tristan Braud, Pan Hui and Lin Wang
\thanks{Corresponding author: Pengyuan Zhou (pyzhou@ustc.edu.cn) and Lin Wang (linwang@ust.hk)}
}


\maketitle
\begin{abstract}
Since 2021, the term ``Metaverse'' has been the most popular one, garnering a lot of interest. Because of its contained environment and built-in computing and networking capabilities, a modern car makes an intriguing location to host its own little metaverse. Additionally, the travellers don't have much to do to pass the time while traveling, making them ideal customers for immersive services. Vetaverse (Vehicular-Metaverse), which we define as the future continuum between vehicular industries and Metaverse, is envisioned as a blended immersive realm that scales up to cities and countries, as digital twins of the intelligent \underline{T}ransportation \underline{S}ystems, referred to as ``TS-Metaverse'', as well as customized XR services inside each \underline{I}ndividual \underline{V}ehicle, referred to as ``IV-Metaverse''. The two subcategories serve fundamentally different purposes, namely long-term interconnection, maintenance, monitoring, and management on scale for large transportation systems (TS), and personalized, private, and immersive infotainment services (IV). By outlining the framework of Vetaverse and examining important enabler technologies, we reveal this impending trend. Additionally, we examine unresolved issues and potential routes for future study while highlighting some intriguing Vetaverse services.
\end{abstract}


\section{Introduction}

Autonomous vehicles have been an engineering goal since the early days of the automotive industry. First attempts at remote-controlled vehicles date back to the 1920s, where a Milwaukee car distributor demonstrated a ``phantom car'' controlled by telegraph keys~\cite{sentinel1926phantom}.
However, there has not been significant breakthroughs until the last decade, enabled by the rapid progress of artificial intelligence (AI).
Mevertheless, self-driving automobiles are at most reaching the ``eyes off'' level defined by the Society of Automotive Engineers (SAE)\footnote{Society of Automotive Engineers (SAE) categorizes autonomous vehicles into six levels: L0 (no automation), L1 (hands-on/shared control), L2 (hands off), L3 (eyes off), L4 (mind off), and L5 (steering wheel optional)~\cite{sae2014automated}} and still require the driver to take back the control if necessary. Vehicles thus still require significant development to achieve full automation. As a result, academia and the automotive industry are under a lot of pressure to stay ahead and come up with new features to keep the field moving forward. 

Among such new features, XR (Extended Reality)  technology has been recently gaining a lot of attention from the automotive industry.  XR is an umbrella term for any technology that alters our perception of the physical world by adding digital elements, including augmented reality (AR), mixed reality (MR) and virtual reality (VR).
XR applications in vehicular environments were initially focused on training. The immersive capabilities of XR allow to develop simulation environments that can assist operators in mastering driving complex vehicles regardless of their location. Companies such as Boeing are already experimenting with virtual training\footnote{\url{https://www.boeingfutureu.com/virtual-experiences/overview}} in an attempt to reduce costs. XR has also been applied to vehicular maintenance. Porshe technicians can use AR glasses that overlays the blueprints onto vehicles and facilitate remote help by specialists, reducing service time by 40\%~\cite{porsche}. Finally, the automotive industry has also experimented with XR at the customer-side, whether in terms of in-car entertainment or to visualize car models in immersive environments. Despite such a strong industry engagement in XR for vehicular applications, academic research remains limited. We have found only 207 papers related to AR/VR vehicular applications, and only 13 on AR/VR and transportation. In comparison to the ambitious moves in the industry and the keen interest of academia for the fields of AR and VR, XR $\&$ vehicle/transportation research is still in its infancy.

Since 2021, the interest in XR technologies has been reinforced by Facebook's announcements relative to the Metaverse, a cyber-physical space that blends physical and digital elements in an immersive manner~\cite{Lee2021AllON}. XR is essential to the current vision of the Metaverse as its primary access medium.
Implementing a Metaverse-like immersive experience within vehicles, dubbed the \textbf{Vetaverse}, appears to be a potential future direction for intra- and inter- vehicular interactions. In addition to the XR applications mentioned above, the Metaverse's pervasive ambitions  will force us to rethink the vehicular ecosystem at a fundamental level. Some pioneer companies have already approached this problem for vehicle production,.  BMW and Nvidia teamed up to build a digital-twin-based virtual factory to reduce production planning~\cite{nvidiabmw}. Beyond the factory, Intelligent transportation systems (ITS) also raise opportunities in the coming Vetaverse era. Nvidia is currently experimenting with city-scale digital twins in South Korea. they leverage data from hundreds of cameras to simulate and optimize a network of roadways~\cite{nvidiakorea}. We believe that the Metaverse is likely to be a substantial trend in the coming years. Accordingly, we conduct this comprehensive survey on the intersection of Metaverse, vehicles, and ITS as a timely guide to the academia and industry. 

\begin{table*}[t!]
\centering
\caption{Comparison of this work with related surveys.}
\begin{tabular}{|llllllllll|}
\hline
\multicolumn{1}{|c|}{Reference} & \multicolumn{1}{c|}{Year} & \multicolumn{1}{c|}{IV appl.} & \multicolumn{1}{c|}{TS appl.} & \multicolumn{1}{c|}{XR} & \multicolumn{1}{c|}{Network} & \multicolumn{1}{c|}{AI} & \multicolumn{1}{c|}{Context} & \multicolumn{1}{c|}{Others}  & Remarks \\ \hline
\multicolumn{10}{|c|}{Surveys on supporting vehicular technologies}  \\ \hline

\multicolumn{1}{|c|}{Lu et al.~\cite{9435134}}     
& \multicolumn{1}{c|}{2022} & \multicolumn{1}{c|}{\cellcolor{green}\checkmark}              & \multicolumn{1}{c|}{\cellcolor{green}\checkmark}      & \multicolumn{1}{c|}{\cellcolor{red}X}   & \multicolumn{1}{c|}{\cellcolor{red}X}   & \multicolumn{1}{c|}{\cellcolor{red}X}   & \multicolumn{1}{c|}{\cellcolor{green}\checkmark}   &  \multicolumn{1}{c|}{\cellcolor{red}X}   &   Localization     \\ \hline

\multicolumn{1}{|c|}{Alalewi et al.~\cite{9497103} }
& \multicolumn{1}{c|}{2021} & \multicolumn{1}{c|}{\cellcolor{green}\checkmark}              & \multicolumn{1}{c|}{\cellcolor{green}\checkmark}              & \multicolumn{1}{c|}{\cellcolor{red}X}   & \multicolumn{1}{c|}{\cellcolor{green}\checkmark}   & \multicolumn{1}{c|}{\cellcolor{red}X}    & \multicolumn{1}{c|}{\cellcolor{red}X}  & \multicolumn{1}{c|}{\cellcolor{red}X}   &    5G and V2X     \\ \hline
\multicolumn{1}{|c|}{Alhilal et al.~\cite{alhilal2022streesmart5G} }     & \multicolumn{1}{c|}{ 2022 }      & 
\multicolumn{1}{c|}{\cellcolor{green}\checkmark}              & \multicolumn{1}{c|}{\cellcolor{green}\checkmark}              & \multicolumn{1}{c|}{\cellcolor{yellow}L}   & \multicolumn{1}{c|}{\cellcolor{green}\checkmark}   & \multicolumn{1}{c|}{\cellcolor{green}\checkmark}    & \multicolumn{1}{c|}{\cellcolor{red}X}  & \multicolumn{1}{c|}{\cellcolor{red}X}   &    5G, V2X, and vehicular applications     \\ \hline

\multicolumn{1}{|c|}{Balkus et al.~\cite{9706268}}          & \multicolumn{1}{c|}{2022} &  \multicolumn{1}{c|}{\cellcolor{green}\checkmark}              & \multicolumn{1}{c|}{\cellcolor{green}\checkmark}              & \multicolumn{1}{c|}{\cellcolor{red}X}   & \multicolumn{1}{c|}{\cellcolor{green}\checkmark}   & \multicolumn{1}{c|}{\cellcolor{green}\checkmark}     & \multicolumn{1}{c|}{\cellcolor{red}X} &  \multicolumn{1}{c|}{\cellcolor{red}X}   &   Collaborative machine learning via V2X.     \\ \hline
\multicolumn{1}{|c|}{Jiang et al.~\cite{jiang2021resource}}    & \multicolumn{1}{c|}{2022}       & \multicolumn{1}{c|}{\cellcolor{green}\checkmark}              & \multicolumn{1}{c|}{\cellcolor{red}X}              & \multicolumn{1}{c|}{\cellcolor{red}X}   & \multicolumn{1}{c|}{\cellcolor{green}\checkmark}   & \multicolumn{1}{c|}{\cellcolor{green}\checkmark}  & \multicolumn{1}{c|}{\cellcolor{red}X}       &  \multicolumn{1}{c|}{\cellcolor{red}X}   &  Video streaming over vehicular networks.    \\ \hline
\multicolumn{1}{|c|}{Fu et al.~\cite{fu2021survey}}      & \multicolumn{1}{c|}{2021}     & \multicolumn{1}{c|}{\cellcolor{green}\checkmark}              & \multicolumn{1}{c|}{\cellcolor{green}\checkmark}              & \multicolumn{1}{c|}{\cellcolor{red}X}   & \multicolumn{1}{c|}{\cellcolor{green}\checkmark}   & \multicolumn{1}{c|}{\cellcolor{green}\checkmark}    & \multicolumn{1}{c|}{\cellcolor{yellow}L}  &   \multicolumn{1}{c|}{\cellcolor{green}\checkmark}   &    Driving safety based on AI and V2X.   \\ \hline
\multicolumn{1}{|c|}{Tang et al.~\cite{tang2021comprehensive}}        & \multicolumn{1}{c|}{2021}   & \multicolumn{1}{c|}{\cellcolor{yellow}L}              & \multicolumn{1}{c|}{\cellcolor{green}\checkmark}              & \multicolumn{1}{c|}{\cellcolor{red}X}   & \multicolumn{1}{c|}{\cellcolor{green}\checkmark}   & \multicolumn{1}{c|}{\cellcolor{green}\checkmark}    & \multicolumn{1}{c|}{\cellcolor{red}X}  &        \multicolumn{1}{c|}{\cellcolor{red}X}   &   Machine learning for vehicular networks. \\\hline
\multicolumn{1}{|c|}{Hu et al.~\cite{hu2022review} }         & \multicolumn{1}{c|}{2022}  & \multicolumn{1}{c|}{\cellcolor{green}\checkmark}              & \multicolumn{1}{c|}{\cellcolor{yellow}L}              & \multicolumn{1}{c|}{\cellcolor{green}\checkmark}   & \multicolumn{1}{c|}{\cellcolor{red}X}   & \multicolumn{1}{c|}{\cellcolor{green}\checkmark}  & \multicolumn{1}{c|}{\cellcolor{green}\checkmark}    &       \multicolumn{1}{c|}{\cellcolor{red}X}   &  Driver digital twin. \\ \hline
\multicolumn{1}{|c|}{Riegler et al.~\cite{riegler2020research}}   & \multicolumn{1}{c|}{ 2020 }        & \multicolumn{1}{c|}{\cellcolor{green}\checkmark}              & \multicolumn{1}{c|}{\cellcolor{red}X}              & \multicolumn{1}{c|}{\cellcolor{green}\checkmark}   & \multicolumn{1}{c|}{\cellcolor{red}X}   & \multicolumn{1}{c|}{\cellcolor{red}X}    & \multicolumn{1}{c|}{\cellcolor{red}X}  &       \multicolumn{1}{c|}{\cellcolor{red}X}   &  Research challenges and agenda for in-car MR. \\ \hline
\multicolumn{1}{|c|}{Tran et al.~\cite{tran2021review}}      & \multicolumn{1}{c|}{ 2021 }     & \multicolumn{1}{c|}{\cellcolor{green}\checkmark}              & \multicolumn{1}{c|}{\cellcolor{red}X}              & \multicolumn{1}{c|}{\cellcolor{green}\checkmark}   & \multicolumn{1}{c|}{\cellcolor{red}X}   & \multicolumn{1}{c|}{\cellcolor{red}X}    & \multicolumn{1}{c|}{\cellcolor{green}\checkmark}  &      \multicolumn{1}{c|}{\cellcolor{green}\checkmark}   &  VR-based vehicle-pedestrian interaction study.  \\ \hline
\multicolumn{10}{|c|}{Surveys on Metaverse}                                                                                                                                                     \\ \hline
\multicolumn{1}{|c|}{Lee et al.~\cite{Lee2021AllON}}     & \multicolumn{1}{c|}{2021}      & \multicolumn{1}{c|}{\cellcolor{green}\checkmark}              & \multicolumn{1}{c|}{\cellcolor{green}\checkmark}              & \multicolumn{1}{c|}{\cellcolor{green}\checkmark}   & \multicolumn{1}{c|}{\cellcolor{yellow}L}   & \multicolumn{1}{c|}{\cellcolor{green}\checkmark}    & \multicolumn{1}{c|}{\cellcolor{red}X}  &   \multicolumn{1}{c|}{\cellcolor{red}X}   & Key techs and ecosystem of Metaverse.      \\ \hline
\multicolumn{1}{|c|}{Lee et al.~\cite{lee2021creators}}        & \multicolumn{1}{c|}{ 2021 }   & \multicolumn{1}{c|}{\cellcolor{green}\checkmark}              & \multicolumn{1}{c|}{\cellcolor{yellow}L}              & \multicolumn{1}{c|}{\cellcolor{green}\checkmark}   & \multicolumn{1}{c|}{\cellcolor{red}X}   & \multicolumn{1}{c|}{\cellcolor{green}\checkmark}     & \multicolumn{1}{c|}{\cellcolor{red}X} &      \multicolumn{1}{c|}{\cellcolor{green}\checkmark}   &  Computational arts for Metaverse.  \\ \hline
\multicolumn{1}{|c|}{Wang et al.~\cite{wang2022survey}}       & \multicolumn{1}{c|}{2022 }    & \multicolumn{1}{c|}{\cellcolor{green}\checkmark}              & \multicolumn{1}{c|}{\cellcolor{green}\checkmark}              & \multicolumn{1}{c|}{\cellcolor{yellow}L}   & \multicolumn{1}{c|}{\cellcolor{green}\checkmark}   & \multicolumn{1}{c|}{\cellcolor{green}\checkmark}   & \multicolumn{1}{c|}{\cellcolor{red}X}   &       \multicolumn{1}{c|}{\cellcolor{green}\checkmark}   & Security and Privacy for Metaverse.   \\ \hline
\multicolumn{1}{|c|}{Ning et al.~\cite{ning2021survey}}      & \multicolumn{1}{c|}{2021 }     & \multicolumn{1}{c|}{\cellcolor{yellow}L}              & \multicolumn{1}{c|}{\cellcolor{yellow}L}              & \multicolumn{1}{c|}{\cellcolor{yellow}L}   & \multicolumn{1}{c|}{\cellcolor{green}\checkmark}   & \multicolumn{1}{c|}{\cellcolor{yellow}L}    & \multicolumn{1}{c|}{\cellcolor{yellow}L}  &       \multicolumn{1}{c|}{\cellcolor{green}\checkmark}   & Techs and Apps for Metaverse.  \\ \hline
\multicolumn{1}{|c|}{Yang et al.~\cite{yang2022fusing}}        & \multicolumn{1}{c|}{ 2022 }   & \multicolumn{1}{c|}{\cellcolor{yellow}L}              & \multicolumn{1}{c|}{\cellcolor{yellow}L}              & \multicolumn{1}{c|}{\cellcolor{red}X}   & \multicolumn{1}{c|}{\cellcolor{green}\checkmark}   & \multicolumn{1}{c|}{\cellcolor{green}\checkmark}    & \multicolumn{1}{c|}{\cellcolor{red}X}  &        \multicolumn{1}{c|}{\cellcolor{green}\checkmark}   & Blockchain and AI for Metaverse. \\ \hline
\multicolumn{10}{|c|}{Surveys on vehicles and Metaverse }                                                                                                                                                  \\ \hline

\multicolumn{1}{|c|}{Our work}    & \multicolumn{1}{c|}{2022}       & \multicolumn{1}{c|}{\cellcolor{green}\checkmark}              & \multicolumn{1}{c|}{\cellcolor{green}\checkmark}              & \multicolumn{1}{c|}{\cellcolor{green}\checkmark}   & \multicolumn{1}{c|}{\cellcolor{green}\checkmark}   & \multicolumn{1}{c|}{\cellcolor{green}\checkmark}   & \multicolumn{1}{c|}{\cellcolor{green}\checkmark}   &   \multicolumn{1}{c|}{\cellcolor{green}\checkmark}   &  \makecell{ Definition of Vetaverse with \\ key enabler techs and envisioned services.}     \\ \hline
\end{tabular}
\caption*{ \fbox{\colorbox{green}{\checkmark} Strong Relevance \colorbox{yellow}{L} Limited Relevance \colorbox{red}{X} Irrelevant}}
\end{table*}

It is important to note that the Metaverse is a complex system that relies on a multitude of supporting technologies. The complexity increases with the amount of interaction between physical elements and their digital counterpart. For instance, accurately developping a digital twin of a vehicle and its driver in real-time requires pervasive sensing and interpretation, both inside-out and outside-in. 
In the vehicle, the on-board unity system (OBU) needs to seamlessly integrate a number of modules, such as sensors, context analysis, information filtering, personalized data mining, and XR rendering, with robust and real-time inter-module data transmissions and substantial processing capabilities. However, vehicles only have a limited vision of the road conditions that needs to be supplemented by other entities  in the urban ecosystem, including other vehicles and roadside units (RSUs)). 
These entities shall communicate with efficient networking technologies and protocols such as C-V2X, DSRC, and 5G/6G to address the real-time constraints of vehicular safety. 
Edge and cloud computing allow aggregating data from multiple road entities, offload some computationally-expensive operations, and process city-scale data into models that can be easily understood by human operators. Data is processed through widespread application of AI techniques, including computer vision (CV) to identify and track objects, and rebuild digital models of road elements.

It would be too long to cover each technology in great details. Therefore, in this work, we present a high-level umbrella vision of existing research related to building such a Vetaverse, with more focus on the primary enabler technologies. These technologies can function in combination with others to provide comprehensive immersive experiences. Next, we examine the existing survey in related directions for a better understanding of our work's position.  



\subsection{Related Surveys and Selection of Articles}
This subsection briefly outlines previous relevant surveys in various domains, i.e., XR for vehicles/transportation systems. 
As early as 2008, Zimmermann~\cite{zimmermann2008virtual} gave an outlook on the use of VR in the automotive industry, which discusses VR-aided design in terms of concerns, concepts, use cases, and economic aspects.
Lawson et al.~\cite{lawson2016future} described an interview with 11 staff from Jaguar Land Rover to analyze the improvement brought by VR in product development and concluded with a positive view of frequently learning from academic literature for better services.
Similarly, Henriques et al.~\cite{henriques2020automotive} tried to identify the opportunities such as cost reduction and proximity to customers, and limitations such as depth and haptic perception and motion tracking, in the application of VR in automotive product development marketing, but mainly through a literature review.
Tran et al.~\cite{tran2021review} reviewed 31 VR-based studies on vehicle-pedestrian interaction to uncover different factors influencing pedestrian experience and behavior, such as characteristics of vehicle, traffic, and environment, and a set of considerations for developing VR pedestrian simulators.
Riegler et al.~\cite{riegler2020research} summarized an MR research roadmap for automated driving based on the spotted challenges, including accessibility, usability testing, and evaluation criteria learned from their workshop and literature review. 
Hu et al.~\cite{hu2022review} systematically illustrated how to build the so-called driver digital twin with key enabling aspects and related technologies, to better understand both the external states such as distraction activity and drowsiness, as well as the internal states such as intentions, emotions, and trust.


\begin{figure*}[t!]
    \centering
    \includegraphics[width=0.9\textwidth]{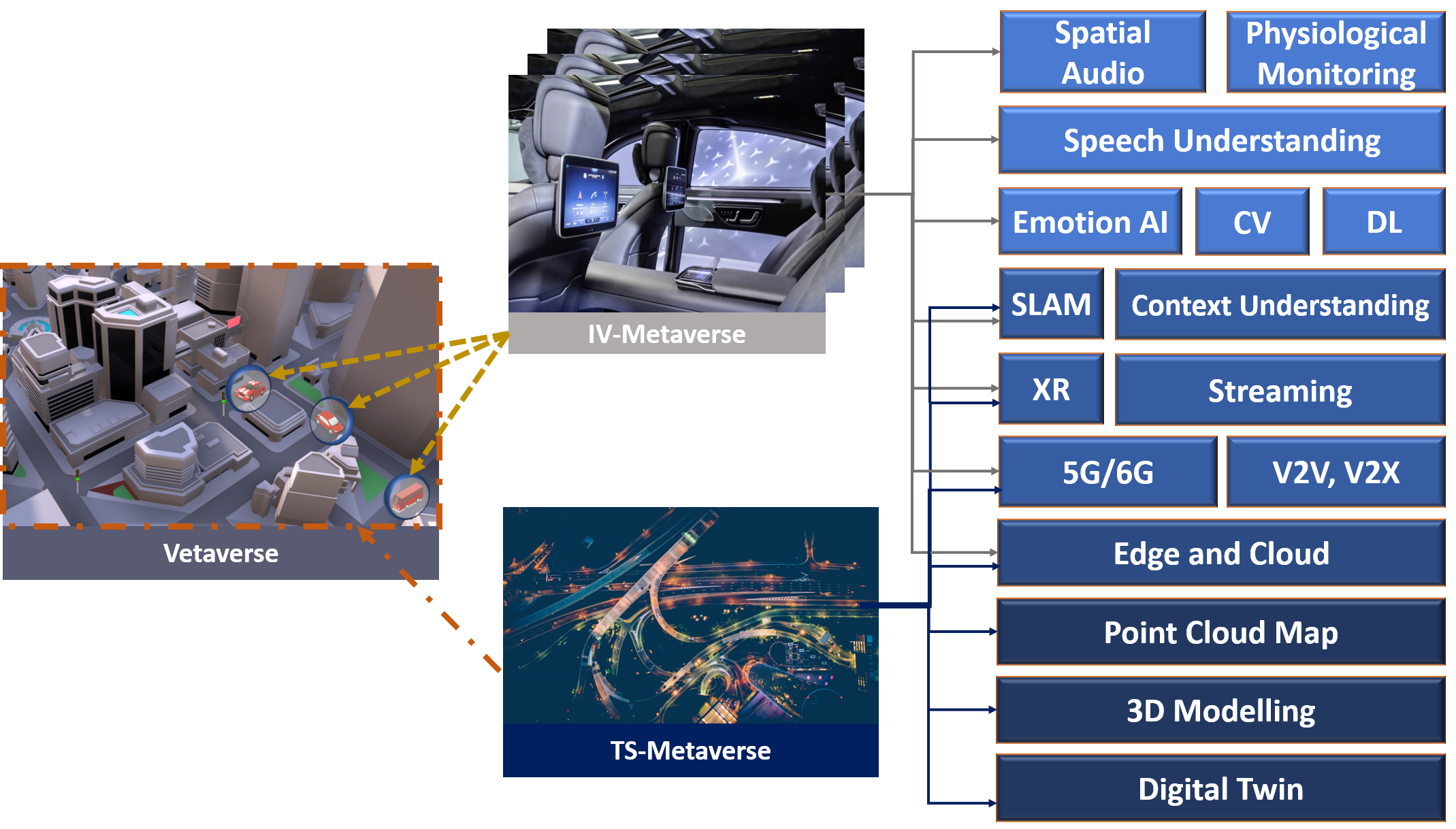}
    \caption{Technology taxonomy of Vetaverse. Each vehicle runs an individual Intra-Vehicle Metaverse consisting of in-car immersive services (see Section~\ref{sec:vehicle}). The numerous IV-Metaverses are connected and mapped into a city-scale digital twin of the transportation system, a.k.a. TS-Metaverse (see Section~\ref{sec:its}). The right side of the figure shows the supporting technologies for the two realms as described in Section~\ref{sec:keytech}. CV: Computer Vision; DL: Deep Learning; SLAM: Simultaneous Localization and Mapping; XR: Extended Reality; V2V: Vehicle-to-Vehicle; V2X: Vehicle-to-Everything.}
    \label{fig:taxonomy}
\end{figure*}

This survey targets a wider view of the interaction between Metaverse and the vehicular field. It covers 200 papers collected from Google Scholar, IEEE Xplore, ACM Digital Library, Elsevier, ScienceDirect, DBLP, etc. The queries combine the following keywords: augmented reality $\&$ vehicle/transportation, virtual reality $\&$ vehicle/transportation, extended reality $\&$ vehicle/transportation, mixed reality $\&$ vehicle/transportation, digital twin, and the keywords related to key technologies discussed in Section~\ref{sec:keytech}. We further complement the articles by adding prominent research featured on the Internet to cover a comprehensive set of important publications in this area. This process was continued until no new articles were found. We have carefully examined the papers and selected the most relevant and important selected articles while filtering out the less relevant ones. The selected papers form the core of this survey, and we have performed continuous updates during the survey writing process to cover papers published since the start of our process.

\subsection{Structure of the survey}
The remainder of this article is organized as follows. Section~\ref{sec:framework} proposes the high-level definitions and overall framework of Vetaverse. Section~\ref{sec:keytech} dives down to key technologies, each of which plays a crucial role in the realm. Section~\ref{sec:its} and Section~\ref{sec:vehicle} give outlooks over the two major parts, Metaverse for ITS, and, Metaverse for in-car services. Finally, we discuss open challenges and future directions in  Section~\ref{sec:discussion} and conclude the work in Section~\ref{sec:conclusion}.

\section{Vetaverse Framework}\label{sec:framework}
\subsection{Overview}
The Metaverse, vehicles, and ITS can interleave in countless ways, e.g., ARHUD for ADAS, in-car VR for infotainment, OBU assistant with avatar, and digital twin of ITS, to name a few. Therefore, accurately defining such an enormous realm is crucial yet challenging. In this work, we propose the skeleton of \textbf{Vetaverse} (Vetaverse), to outline an umbrella of the future interaction between vehicular industries, transportation industries, and the Metaverse. We define Vetaverse, according to whether focusing on intra-vehicle services, in two aspects: \textbf{IV-Metaverse} (intra-vehicle Metaverse) and \textbf{TS-Metaverse} (transportation system Metaverse). Figure~\ref{fig:taxonomy} depicts our proposed taxonomy for the skeleton and technology stack of Vetaverse. The overall ITS is mapped into a huge digital twin, in which each individual vehicle runs its own in-car IV-Metaverse that is linked to other IV-Metaverses in the shared digital twin, composing the TS-Metaverse. IV-Metaverse, serving in-car drivers and passengers, and, TS-Metaverse, serving all road users, have different focuses of services thus requiring different technologies but with some overlaps. Next, we briefly describe the vision of two aspects. We detail the technology stack in Section~\ref{sec:keytech} and envisioned services in Section~\ref{sec:its} and Section~\ref{sec:vehicle}.

\begin{figure*}[t!]
    \centering
    \includegraphics[width=0.9\textwidth]{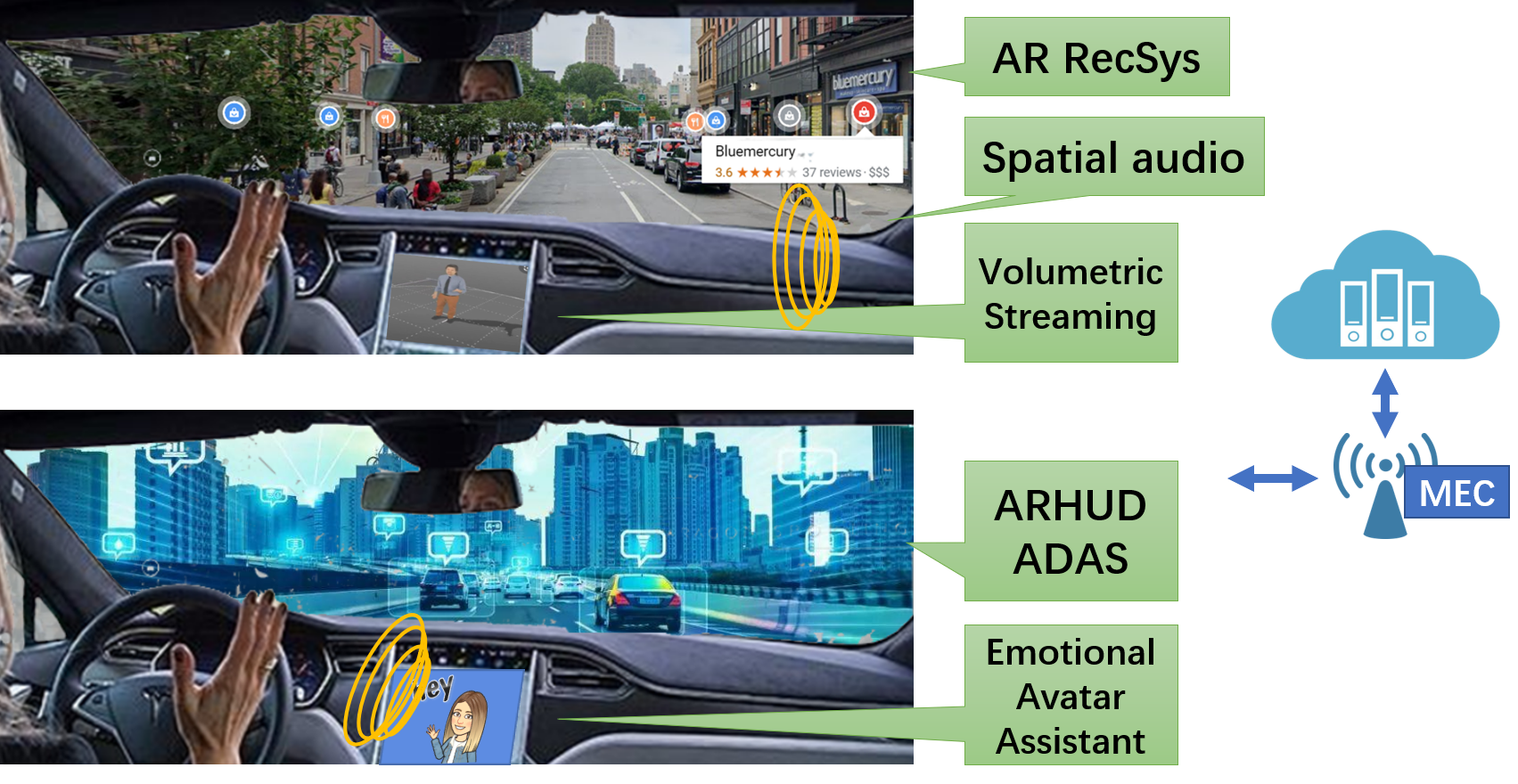}
    \caption{Intra-Vehicle Metaverse services can provide users in the vehicle with immersive experience during commute. For instance, ARHUD can provide AR-based recommendation system for POI discovery and augmented information for driving assistance. Volumetric streaming allows the users enjoy volumetric videos in the car. Spatial audio and emotional AI can improve the understanding and feedback of OBU assistant for more user friendly service experience. }
    \label{fig:IV-Metaverse}
\end{figure*}

\textbf{IV-Metaverse.} focuses on the infotainment services provided to the drivers and passengers through the windshield and side windows to experience a seamless connection between the real and virtual worlds. As depicted in Figure~\ref{fig:IV-Metaverse}, the vehicle provides an immersive Metaverse experience powered by AR, AI, spatial audio, streaming, edge computing, and V2X technologies. Potential services include user-friendly assistant based on emotional AI~\cite{zepf2020driver}, ARHUD-based driving assistance facilitated by V2X and context understanding~\cite{aicp,9163287}, AR-RecSys (recommendation system) according to situation awareness and personalized training, volumetric streaming services optimized by privacy-preserved learning techniques~\cite{gao2022fras}, and realistic spatial audio services~\cite{kari2021soundsride}, etc.

\begin{figure*}[t!]
    \centering
    \includegraphics[width=0.7\textwidth]{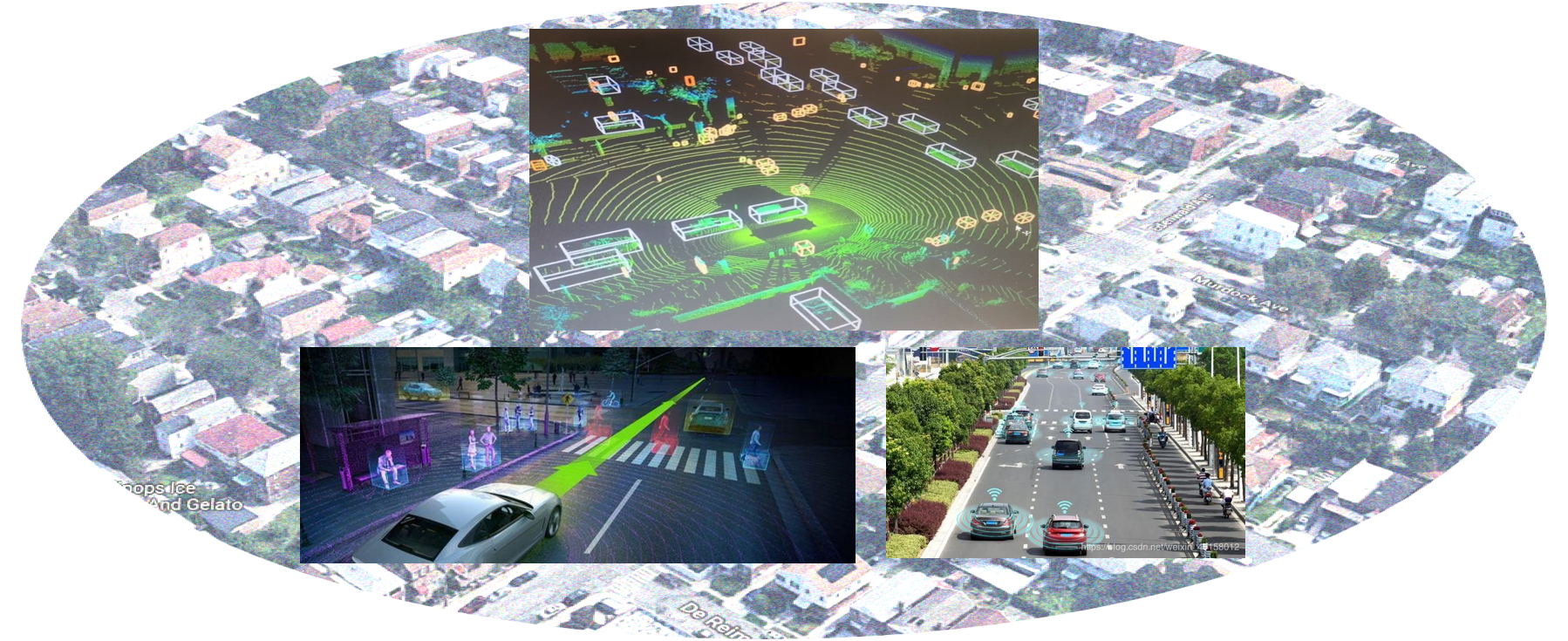}
    \caption{Transportation-System Metaverse. Each vehicle detects its surroundings via sensory data (bottom left of the figure) and communicates the learned context with nearby vehicles via V2V (bottom right) and aggregates it to the cloud via V2X. The central server maintain the TS-Metaverse using the aggregated context data (top of the figure).}
    \label{fig:ts-Metaverse}
\end{figure*}

\textbf{TS-Metaverse.} 
Based on V2X, edge, and cloud techniques, TS-Metaverse provides a universal immersive platform to host the digital twin and shadow of everything on the road for real-time and long-term traffic monitoring and large-scale on-road interactions as depicted in Figure~\ref{fig:ts-Metaverse}. Sensing technologies such as LiDAR enable the situation awareness capability for the vehicles to accurately reconstruct surrounding environments with 3D reconstruction techniques, as shown at the left bottom in the figure. V2X technologies link every machine on the road to exchange critical information, including each vehicle's sensed surroundings in real-time, as shown at the right bottom in the figure. The surrounding environments sensed by the vehicles are connected and mapped into the ``global'' digital twin of the transportation system, collectively built with the smart city sensors, compromising a TS-Metavese that can monitor the transportation in 3D as well as allow intercommunication between the users' IV-Metaverses.

\section{Key Technologies}\label{sec:keytech}
Building the futuristic Metaverse requires a lot of technologies to function integrally. We briefly lay out the key technologies that play crucial roles following the top-down sequence shown in Figure~\ref{fig:taxonomy}.

The most human-centered interaction between a vehicle and the driver/passengers is the intelligent car assistant~\cite{williams2013towards} running in the central control module (CCM) that listens to the audio and reacts with provided services. 
Two fundamental functions to guarantee the QoE of such services are speech understanding~\cite{deng2013recent} and emotion comprehension~\cite{zepf2020driver}, which require speech recognition and synthesis, physiological monitoring, NLP analysis, and emotional AI, etc. Speech understanding allows the assistant to understand the intention of the driver's command and provide the demanded services. Emotional AI allows the assistant to understand the inertial mood of the driver and give proper reactions, such as responding in proper tones or showing the avatar in proper expressions. To further augment the assistant's service, spatial audio can be employed to control the audio effect and angle for an immersive experience.

The next major driver-vehicle interaction is the driving assistant functions that provide context awareness and action recommendations. For instance, CV, DL (deep learning), and SLAM capacities allow the recognition of the surrounding dangers or POI (point of interest)~\cite{khurshid2020scene,hbali2016face,mur2017orb}, which can be displayed to the driver via AR display such as ARHUD~\cite{arve}, for better context understanding. Users can also enjoy immersive video streaming services powered by advanced networking technologies such as 5G/6G~\cite{gao2022fras}. Nevertheless, given the trend of electric vehicles, battery consumption would become a major concern for most future vehicles in the Vetaverse era since most of the aforementioned immersive services incorporate complex computation tasks that drain the battery. V2V/V2X, Edge, and Cloud technologies can all assist by orchestrating resource and task allocations between vehicles, edge, and cloud for optimized offloading services to reduce vehicle computation burden~\cite{alhilal2021cad3,chang20226g,9163287,zhou20215g}. 

Built on a point cloud map, TS-Metaverse can show how ITS works in the real world in almost real-time because it constantly gets updated data from city-scale distributed sensors. TS-Metaverse can enable several exciting futuristic approaches: 1) Near-real-time 3D traffic and ITS monitoring. Current map services like Google Maps can provide 3D navigation from the user's point of view to explore points of interest (POI) using the Google Earth service. However, none of the businesses in the market can provide a near real-time 3D digital twin of the real-world ITS and traffic on large scale. Sensory data fusion and 3D modeling can enable a twin-based simulation environment for autonomous driving by generating realistic data that can be fed into simulated cameras, LiDAR, and other sensors \cite{airsim2017fsr,Dosovitskiy17}. Advanced communication and control technologies make it possible for digital twins and physical objects to share data and control each other in real-time, realizing the digital shadow of ITS.

Next, we describe the key enabler technologies in detail.

\subsection{Artificial Intelligence}
AI has been broadly deployed in vehicles and transportation systems, in terms of data mining, decision-making, context understanding, computer vision tasks, etc. Notably, these tasks are often interwoven and required interleaved AI operations. For instance, context understanding normally requires deep learning for analyzing sensory data, computer vision algorithms for processing camera-captured images, and 3D reconstruction algorithms to further comprehend the 3D geometry and structure of objects and scenes for a better understanding of the surroundings. More specifically, the aforementioned tasks require machine learning algorithms for scene classification~\cite{khurshid2020scene}, object detection~\cite{liu2019edge}, semantic segmentation~\cite{wang2021evdistill}, depth estimation~\cite{prokopetc2019towards}, pose estimation~\cite{li2021exploring}, image restoration \cite{liang2021swinir} and augmentation~\cite{li2021low} etc., which demands a variety of neural networks such as Convolutional Neural Network~(CNN), Recurrent Neural Network~(RNN), Transformer, and Deep Neural Network~(DNN) etc. Please refer to Section~\ref{subsec:context} and  Section~\ref{subse:3dreconstruct} for more details of AI usage in context understanding and computer vision tasks. Reinforcement learning has also been employed for large-scale intelligent traffic light control~\cite{zhou2019erl}, which can control the lights according to real-time traffic conditions. Decentralized-learning-based traffic light systems can collect and aggregate data to the ITS digital twin (TS-Metaverse) in real-time, thanks to edge and cloud computing~\cite{zhou2020drle}.

In Vetaverse, AI can play a more important role. For example, emotion AI can drive the development of user-friendly car assistant that can understand and react to the driver's and passengers' emotions, which requires several AI-enabled technologies~(see Section~\ref{subsec:emotion} for details). Moreover, deep learning techniques can be applied for optimized resource allocation and task assignment~\cite{9706268}, which is crucial for computation-intensive and latency-sensitive metaverse applications. For instance, federated learning can be applied for adaptive bitrate encoding for seamless volumetric in-car streaming services while protecting user privacy~\cite{gao2022fras}.

\subsection{Speech Understanding} 
Since connection or communication in virtual spaces is becoming an essential part of future, being able to understand speech becomes very important for AR/MR-based interaction systems. Unlike context understanding, speech understanding facilitates the understanding or communication with objects and avatars without requiring visual attention or physical manipulation. Speech understanding is also used to lower the risk of car accidents by letting drivers give commands with their voices. Speech recognition makes it possible for drivers to directly control devices without having to move their eyes or hands. Speech enhancement is used to improve the voice of vehicle noise. This makes it easier to get rid of the noise from the vehicle and improves the rate of speech recognition in the vehicle. Moreover, speech emotion recognition~\cite{tan2021speech} can perceive and analyze drivers' emotional expressions in vehicles, and then vehicles can perceive drivers' emotions and respond accordingly. This makes it possible to enhance the interactive relationship between drivers and vehicles. Speech interaction has recently been exploited for MR-based maintenance to help users control the workflow when engineers’ hands are busy with equipment~\cite{siyaev2021towards}. Similarly, \cite{SiyaevJ21} proposed the Aircraft Maintenance Metaverse for aircraft maintenance training and education of Boeing-737 and designed a speech understanding module to control the virtual environment effectively. In summary, speech understanding is significantly crucial for guaranteeing the achievement of the Vetaverse. Speech understanding in Vetaverse aims to: 1) remove the vehicle noise; 2) improve communication between avatars; 3) operate devices in Vetaverse accurately. GGS-AR~\cite{WangWYL21} proposes the speech-based interaction through verbal command, enabling users to interact with AR objects. Furthermore, Lamberti et al.~\cite{LambertiMPPS17} elaborated a framework that can automatically generate speech-based interfaces to control VR and AR applications on wearable devices. For AR applications, Williams et al.~\cite{WilliamsGO22} proposed to combine gesture and speech for object manipulation, where speech is adopted for creating new objects. 

\begin{figure}[t!]
    \centering
    \includegraphics[width=\columnwidth]{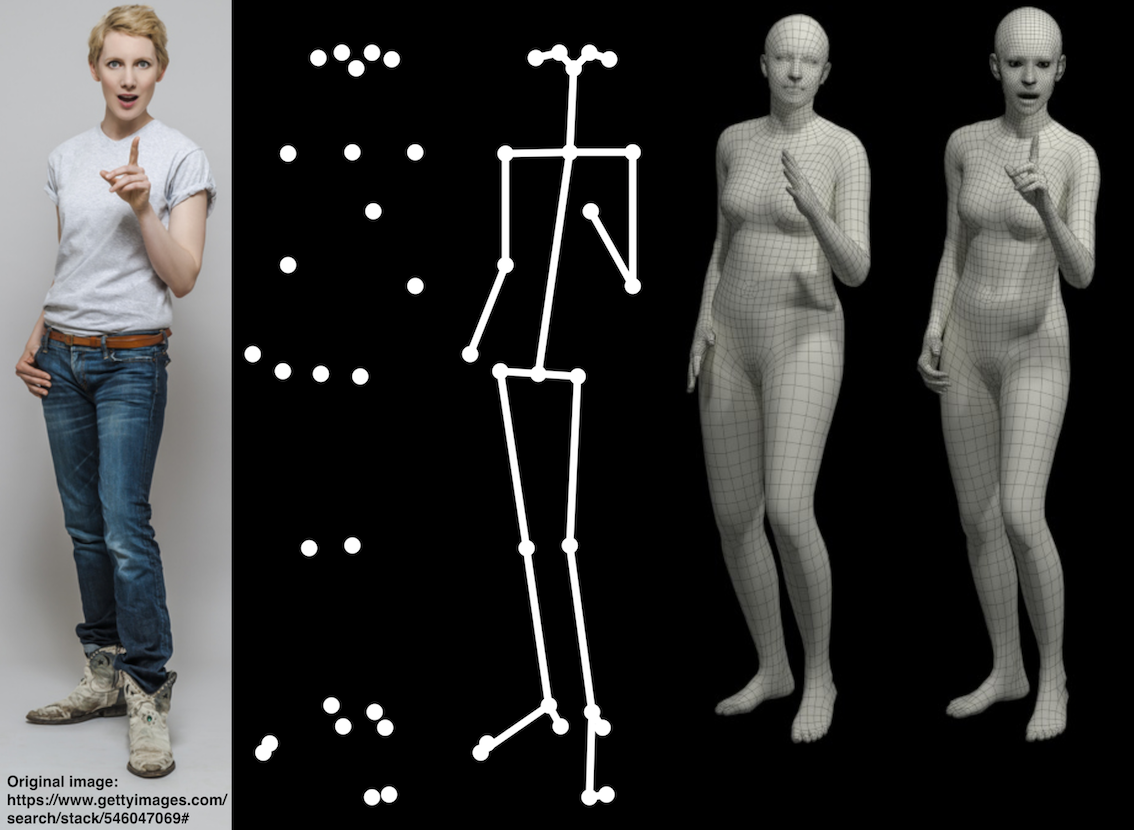}
    \caption{This figure shows the human key points detection and shape reconstruction results.}
    \label{fig:human-keypoint-and-shape}
\end{figure}
\subsection{Human Key Point Detection and Shape Reconstruction}


The goal of human keypoints detection is to locate the positions of keypoints of the human body, such as the head, shoulders, and knees~\cite{fang2017rmpe}, to facilitate driver and passenger monitoring. 
Keypoints detection includes 2D keypoint detection~\cite{8765346,simon2017hand,cao2017realtime,wei2016cpm} and 3D keypoint detection~\cite{cheng2019occlusion,guler2019holopose}. 
The coordinates of 2D keypoint positioning are the pixel coordinates in the image. 
The coordinates of the 3D keypoint are the relative coordinates of the human body key points in three-dimensional space.
Human shape reconstruction~\cite{kolotouros2019convolutional} is highly related to  human keypoint detection.
It focuses on reconstructing the 3D shape of the human body which contains more information than keypoint locations.
Human keypoint detection methods can be divided into two categories: bottom-up~\cite{8765346,cao2017realtime} and top-down~\cite{xiu2018poseflow,li2021hybrik,dang2019deep}. 
In general, top-down methods have high accuracy but require more computation, while bottom-up methods are faster and easier to deploy.

\subsection{Eye Tracking}

\subsection{Physiological parameter monitoring}

The multi-parameters of the driver's physical condition are important indicators to measure the healthy driving state. Fatigue driving, drunken driving, and sudden illness during driving will cause changes in these parameters, and they are also one of the main causes of traffic accidents. Therefore, it is necessary for Vetaverse to monitor the values of these parameters and opportunely adjust the vehicle's driving state to make prevention. The existing life feature extraction work has achieved non-interference extraction of driver's heart rate based on multiple imaging ~\cite{Huang2021AHR, Selvaraju2022UnobtrusiveHR}, non-contact detection of ECG ~\cite{Leicht2017ClosedLoopCO}, detection of blood alcohol concentration ~\cite{Sakairi2012WaterClusterDetectingBS} and simultaneous extraction of heart rate and respiratory rate based on radar ~\cite{Schires2018VitalSM, Chen2022DetectDUIAI}. Some indirect parameters, such as cognitive load, can also be detected by detecting pupil size ~\cite{Fridman2018CognitiveLE}. When the detected parameters are abnormal, Vetaverse is supposed to inform relatives, friends, or traffic agencies of the driver's dangerous driving circumstance and propose treatment or relief programs. For instance, VR has been proven to relieve anxiety and pain, and reduce heart rate and blood pressure ~\cite{Wiederhold1998ARO}.


\subsection{Emotion recognition}\label{subsec:emotion} 

Emotion recognition has the potential to be applied to natural language processing (NLP) assistant, digital twin $\&$ avatar reconstruction, and wireless emotion recognition.

\subsubsection{NLP Assistant} 

NLP-based emotion recognition can help non-player character(NPC) or AI assistants in the Vetaverse understand human feelings and needs for a better human-machine interaction experience. With input signals such as text, speech, or conversation, researchers have already developed various emotion recognition methods \cite{deng2021survey,poria2019emotion,khalil2019speech}. 

However, there are still many problems to be solved before the neural network models can fully understand how people feel and help people in the Vetaverse. For example, the textual emotion recognition approaches suffer from a shortage of high-quality datasets, fuzzy emotional boundaries and incomplete emotional information in the textual expressions \cite{deng2021survey}. Existing speech-based emotion recognition methods have limitations such as large architectures and lower efficiency for the temporally varying input data, implying that they are not robust enough for dynamic scenarios in the Vetaverse. The conversation-based emotion recognition method faces the challenges of speaker-specific modeling and multiparty conversation, which is not user-friendly enough when there are multiple users in the Vetaverse \cite{khalil2019speech}. In conclusion, for Vetaverse to understand how users feel, it needs better, lighter NLP models with stable efficiency and high accuracy. 

\subsubsection{Digital Twin $\&$ Avatar Reconstruction}

Avatar emotion reconstruction is more appealing than the traditional communication methods such as texting, phone calls, and video meetings because users can use dynamic avatars to represent different emotional states and interact with the Vetaverse participants\cite{del2021interpersonal}. According to \cite{del2021interpersonal}, dynamic facial expressions rendered by avatars can generate a more immersive emotional experience for users. Existing emotion recognition and reconstruction approaches can help reconstruct the avatars’ emotions according to the users’ facial expressions, which makes communication between Vetaverse users more understandable. The most commonly used methods for face reconstruction involve statistical model fitting methods, photometric methods, and deep learning-based methods \cite{morales2021survey}. By using emotion recognition, not only can human-shaped avatars express emotions, but also characteristically shaped avatars such as animals or emojis can generate emotions. 

Various sensors can capture different signals for ML-based emotion recognition and avatar emotion reconstruction. \cite{purps2021reconstructing} proposed a method to reconstruct facial expressions of head-mounted display users for avatars by developing a three-stage machine learning pipeline and creating realistic avatars using photogrammetry, 3D modeling, and applied blend-shapes. In addition, the implementation of key functions in the driver digital twins is also inseparable from emotion recognition, 
such as driver distraction detection, driver drowsiness detection, and driver emotional state monitoring and so on \cite{hu2022review}. For example, \cite{ramos2021building} conducted emotion recognition in the digital twin by using a brain-computer interface. \cite{danvevcek2022emoca} proposed a novel deep learning model to reconstruct 3D facial avatars for digital twins from monocular images. \cite{wu2019continuous} proposed a vision-based approach for digital twins, in which facial expressions are identified through analyzing driver head poses and gaze captured by camera sensors. The ability to reconstruct emotions in real-time is vital for the Vetaverse to work and is essential for avatars to communicate better with each other. 

\subsubsection{Safe Driving $\&$ Health Care}

In Vetaverse, it is always essential to ensure passengers’ safety with potential danger warnings and health care. There are mainly two aspects of how emotion recognition could help with the safety and health of Vetaverse users. One is to monitor the driver’s states to warn the driver of the potential danger, especially for vehicles without full automation~\cite{sae2014automated}. The other one is using various signals such as electrocardiograms and facial images to conduct emotion recognition and further monitor drivers’ and passengers’ physical and psychological conditions for healthcare purposes. 

Emotion recognition can be used in driver digital twins to monitor drivers’ conditions, which can help supervise drivers’ emotional conditions and prevent them from dangerous driving behaviour. For instance, \cite{zepf2020driver} designed a driver state monitoring (DSM) system to analyze drivers’ conditions and sound the alarm by emotion recognition based on monitoring drivers’ faces. Biophysiological signals captured by wearable devices can also help to estimate if the driver is under abnormal emotional conditions such as stress and anger \cite{zepf2020driver,katsigiannis2017dreamer}. 

In Vetaverse, it is also possible to apply emotion recognition to healthcare. Emotion recognition-based healthcare can be implemented by capturing users’ facial images, speech, or electrocardiograms via cameras and wearable devices. For example, according to \cite{el2019virtual}, it is plausible to apply NLP to VR/AR virtual patient systems to recognize users’ emotions; \cite{hasnul2021electrocardiogram} uses electrocardiograms to estimate if patients are stressed, anguished, or depressed; \cite{dhuheir2021emotion} uses audio-visual signals such as facial images to evaluate patients’ abnormal conditions; \cite{subramanian2022digital} proposed a digital twin model, which is capable of conducting real-time emotion recognition for personalized healthcare using an end-to-end framework.  In this way, the health conditions of users can be monitored.  

For the Vetaverse to operate safely, it is essential to have functions for monitoring the safety and health of users through emotion recognition. However, as people’s faces are usually captured for this purpose, it remains a challenge for human privacy protection in the future \cite{wang2022survey}.

\subsubsection{Wireless emotion recognition}
Emotion recognition through wireless methods has also been explored in recent years. This strategy was first proposed in \cite{zhao2016emotion} to capture emotion information via radio frequency signals containing heartbeat segmentation information reflected off the body. EmoSense, a low-cost solution for emotion sensing combined with off-the-shelf WiFi devices, was proposed in \cite{8422330}. In a smart home scenario, \cite{8105799} combined different IoT devices to capture speech and image signals for emotion recognition.

\subsection{Spatial Audio}
Spatial audio~\cite{rumsey2012spatial}, also referred to as 3D audio or 360 audio, can be produced by stereo speakers, headphones, surround-sound speakers, or speaker-arrays. When playing spatial audio, it changes with the movement of the viewer’s head. The common technique is spatial domain convolution of sound waves based on head-related transfer functions (HRTF)~\cite{so2006low}. HRTF is a sound localization algorithm where a pair of binaural HRTFs can be used to synthesize binaural sounds from specific points in space~\cite{schissler2016efficient}. It describes how sound from a specific point reaches the ear (usually the front of the tympanic membrane). Also, the signal from any point in space to the human ear can be described by a filter that can describe the spatial information. Then it is possible to restore the sound signal from this orientation in space~\cite{gupta2010hrtf}.

Due to its lightweight and real-time operation, spatial audio can bring great convenience to drivers~\cite{dupreSpatialSoundDesign2021}. Spatial audio is a critical component in creating truly immersive AR/MR-based systems~\cite{spagnol2018current}. Spatial audio technologies currently have many different application areas, such as personal entertainment devices, real-time communication systems~\cite{yao2017headphone}, acoustic-based location monitoring~\cite{marentakis2021spatial}, etc. In addition, they are also widely used in transportation applications~\cite{dupreSpatialSoundDesign2021}.

\subsection{Context Understanding}\label{subsec:context}
\subsubsection{Holistic Scene Understanding}
Holistic scene understanding is an indispensable part of achieving the perception of any AR/MR-based interaction systems. It ensures interaction with other objects and avatars in both the virtual and physical worlds. Therefore, scene understanding is the key technique that guarantees the functioning of the Vetaverse. It aims to: 1) understand what kind of scene a vehicle is expected to perceive, e.g., if a vehicle moves along the urban roads, it mostly sees buildings, vehicles, pedestrians, etc; 2) understand what kind of contents or objects are present around an avatar, namely a representative of a person in the physical world; 3) what are the semantic regions of the contents, e.g., buildings and roads; 4) how to deal with unfavorable weather or lighting condition for holistic scene understanding, 
e.g. rain, under-exposure, etc. Holistic scene understanding combines these aspects of a scene for a vehicle to provide a comprehensive understanding of the whole scene in the Vetaverse. In general, holistic scene understanding can be decomposed into four major tasks: scene classification, object detection, image segmentation, and image restoration and enhancement. 
 
\subsubsection{Scene Classification}
Scene classification in Vetaverse aims to classify a scene into a certain scene type (e.g., urban, rural, and seaside) based on the ambient content, objects, and their layouts~\cite{khurshid2020scene}. The core technique of scene classification is to transform a scene image into feature descriptors to recognize the scene. For AR-based Vetaverse, a representative approach for scene classification is dependent on an AR device to capture images first and a CNN-based framework to perform scene classification~\cite{khurshid2020scene,kalkofen2008comprehensible}. For accomplishing the Vetaverse, one inevitable challenge is how to define the complexity of the virtual world so as to perform scene classification. For instance, the objects and contents in the Vetaverse can dynamically vary, and the avatar types can also change according to the driver's preferences. Consequently, these uncertainties cause considerable technical difficulties for scene classification in the Vetaverse system. Moreover, it is expected to perform real-time scene classification in the Vetaverse. However, due to the complexity of the driving environment, it is difficult to deploy computing devices to meet such a requirement.    

\begin{figure}[t!]
    \centering
    \includegraphics[width=\columnwidth]{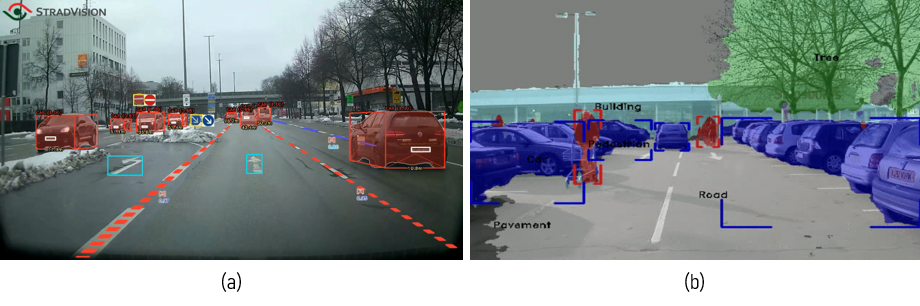}
    \caption{(a) An example of 3D object detection (e.g., vehicles, landmarks) in AR-based scene perception for urban driving (Image source: Stradvision). (b) An example of semantic segmentation in AR for driving scene understanding~\cite{siam2017deep}}
    \label{fig:seg-det}
\end{figure}

 \subsubsection{Object detection}
  Objection detection aims to localize the objects in a scene and identify the class information for each object~\cite{liu2019edge}. Object detection in the Vetaverse can be classified into two categories: detection of specific instances (e.g., faces, markers, and text) and detection of generic categories (e.g., cars and humans). Text detection methods have been broadly studied in XR~\cite{hbali2016face}. These methods have already matured and can be directly applied to achieving the Vetaverse. Face detection has also been studied extensively in recent years, and the methods have been shown to be robust in various recognition scenarios in XR applications, e.g.,~\cite{lu2020exploration,kojic2020user}.
  For example, in VR, face detection is a typical object detection task, while pedestrian detection is a common object detection task in AR for autonomous driving. 

 In a more sophisticated application, AR object recognition aims to attach a 3D model to the physical world~\cite{li2020object}. This requires the object detection algorithms to precisely locate the position of objects and correctly recognize their classes. By placing a 3D virtual object and connecting it with the physical object, users can manipulate and relocate it. AR object detection can help build a richer and more immersive 3D environment in the Vetaverse.  A representative example is StradVision~\footnote{\url{https://www.youtube.com/watch?v=tK7F8yvpiGA}} which subtly leverages AI-based vehicle detection, lane detection, and semantic segmentation algorithms to build an AR-based ADAS system, as shown in Figure~\ref{fig:seg-det}(b).


In the Vetaverse, users are represented as avatars, and multiple avatars can interact with each other. The face detection algorithms need to detect both real faces from the physical world and synthetic faces from the virtual world. Moreover, the occlusion problems, sudden face pose changes, and illumination variations in the Vetaverse can make it more challenging to detect faces in the Vetaverse. Another problem with face detection is the privacy risk. Several research works have studied this problem in AR application~\cite{acquisti2014face}. In the Vetaverse, many users can stay in the 3D immersive environment; hence, privacy in face detection can be more stringent. Future research should consider the robustness of face detection, and better rules or criteria need to be studied.     


 
 
 \subsubsection{Image Segmentation}
Image segmentation aims to categorize an image into different classes based on the per-pixel information~\cite{wang2021evdistill,wang2021dual}. It is considered as one of the fundamental techniques to understand the driving environment in the Vetaverse~\cite{tanzi2021real}. Semantic segmentation is one typical image segmentation task, aiming to efficiently and quickly segment each pixel based on the class information. In recent years, AI-driven semantic segmentation methods have demonstrated promising performance on numerous urban driving datasets of self-driving scenarios. Nonetheless, the XR-based applications for driving usually require the semantic segmentation algorithms to have an inference speed of at least 60 frames per second (fps)~\cite{ko2020novel}. To achieve the Vetaverse, it is pivotal to compress the AI-driven semantic segmentation models and address the problem of heavy computation costs.






\begin{figure}[t!]
    \centering
    \includegraphics[width=.8\columnwidth]{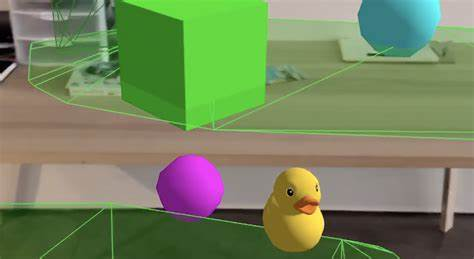}
    \caption{An example of semantic segmentation in AR\protect\footnotemark.}
    \label{fig:segmentation}
\end{figure}
\footnotetext{\url{https://blog.mozvr.com/semantic-placement-in-ar/}}

The early attempts of semantic segmentation mostly utilize the feature tracking algorithms, e.g., SIFT~\cite{schonberger2018semantic}, which aims to segment the pixels based on the classification of the handcrafted features, such as the support vector machine (SVM)~\cite{noble2006support}. These algorithms have been applied to VR~\cite{lin2016virtual} and AR~\cite{schutt2019semantic}. However, these conventional methods suffer from limited segmentation performance. Recent research has explored the potential of CNNs for semantic segmentation. These methods have been successfully applied to AR~\cite{ko2020novel,tanzi2021real}. Some works have shown the capability of semantic segmentation for tackling the occlusion problems in MR~\cite{kido2021assessing,roxas2018occlusion}. However, as image segmentation deals with each pixel, it leads to considerable computation and memory load.

To tackle this problem, recent endeavors have focused on real-time semantic segmentation. Theses methods explore the image crop/resizing~\cite{zhao2018icnet} or efficient network design ~\cite{siam2018rtseg,mehta2018espnet} or model compression~\cite{wang2021knowledge}. Through these techniques, some research works managed to achieve real-time semantic segmentation in MR~\cite{kido2020mobile,gajic2020egocentric,chen2020context}.  

In the Vetaverse, we need more robust and real-time semantic segmentation methods to understand the pixel-wise information in a 3D immersive world, as shown in Figure~\ref{fig:segmentation}. Because of the diversity and complexity of virtual and real objects, contents, and human avatars, more adaptive semantic segmentation methods are required. In particular, in the interlaced Vetaverse world, semantic segmentation algorithms must distinguish virtual object pixels from real ones. In this case, class information may be more complex, and semantic segmentation models may have to deal with previously unseen classes.

\subsubsection{Image Restoration and Enhancement}

Image restoration and enhancement aim to improve image quality in the Vetaverse for more robust scene understanding. The traditional VR wearable prototype or AR display produces poor image quality and contrast due to hardware limitations and other factors. Image restoration has the potential to be useful in revealing more details with higher image resolution for a more pleasant user experience. For example, \cite{kim2022holographic} designs small holographic glasses for VR, aiming to solve the problem of poor image quality with a near-eye display system by employing the HOGD-CITL algorithm. 

The Vetaverse has the potential to speed up the development of autonomous driving with better scene understanding abilities through image restoration and enhancement. On the one hand, higher-quality images can build up users' trust in this new technology and improve drivers’ safety via enhanced context awareness with ``Invisible-to-Visible''(I2V) \cite {lindemann2018supporting,njoku2022prospects} through XR devices. I2V is a technology that can extend drivers' vision ability, as well as for corner case scene understanding, such as blind spots, in adverse weather, at a greater distance, or in poor lighting conditions, where human visual ability is greatly influenced. A nighttime AR system, for example, is intended to improve driving safety at night \cite{ park2015augmented}.  On the other hand, the intelligent perception and decision systems of the vehicle rely heavily on the quality of the collected data. Therefore, it is critical for the vehicle systems to have a better understanding of the environment in order for the intelligent system to make more accurate decisions. Moreover, with the diverse development of deep learning methods in image restoration \cite{liang2021swinir}, techniques such as derain \cite{ wang2019spatial}, defog \cite{ arif2022comprehensive} and image enhancement \cite{li2021low} are potentially useful to solve the challenges of image quality degradation so as to achieve a more robust Vetaverse.

\subsection{Image-based 3D Reconstruction}\label{subse:3dreconstruct} 
In the Vetaverse, drivers and passengers and their digital representatives (i.e., avatars) are connected and interact with each other at the junction of the physical driving environment and the virtual world. Therefore, building such a junction across the physical driving scene and the virtual environment inspires us to deeply understand human behaviors, which empower the action of the avatar and the surroundings of the vehicle. In the physical driving environment, the driver can estimate the overall size and geometry of an object with their eyes and construct a 3D world based on our mental model of what an object looks like. Similarly, the Vetaverse also needs to build up the 3D structure of the driving environment and localize the moving objects. Image-based 3D reconstruction is a hot topic in computer vision and graphics that infers the 3D geometry and structure of objects and scenes (e.g., vehicles) from visual cues such as RGB image~\cite{yang2013image,ghasemieh20223d}.  

\begin{figure}[t!]
    \centering
    \includegraphics[width=.8\columnwidth]{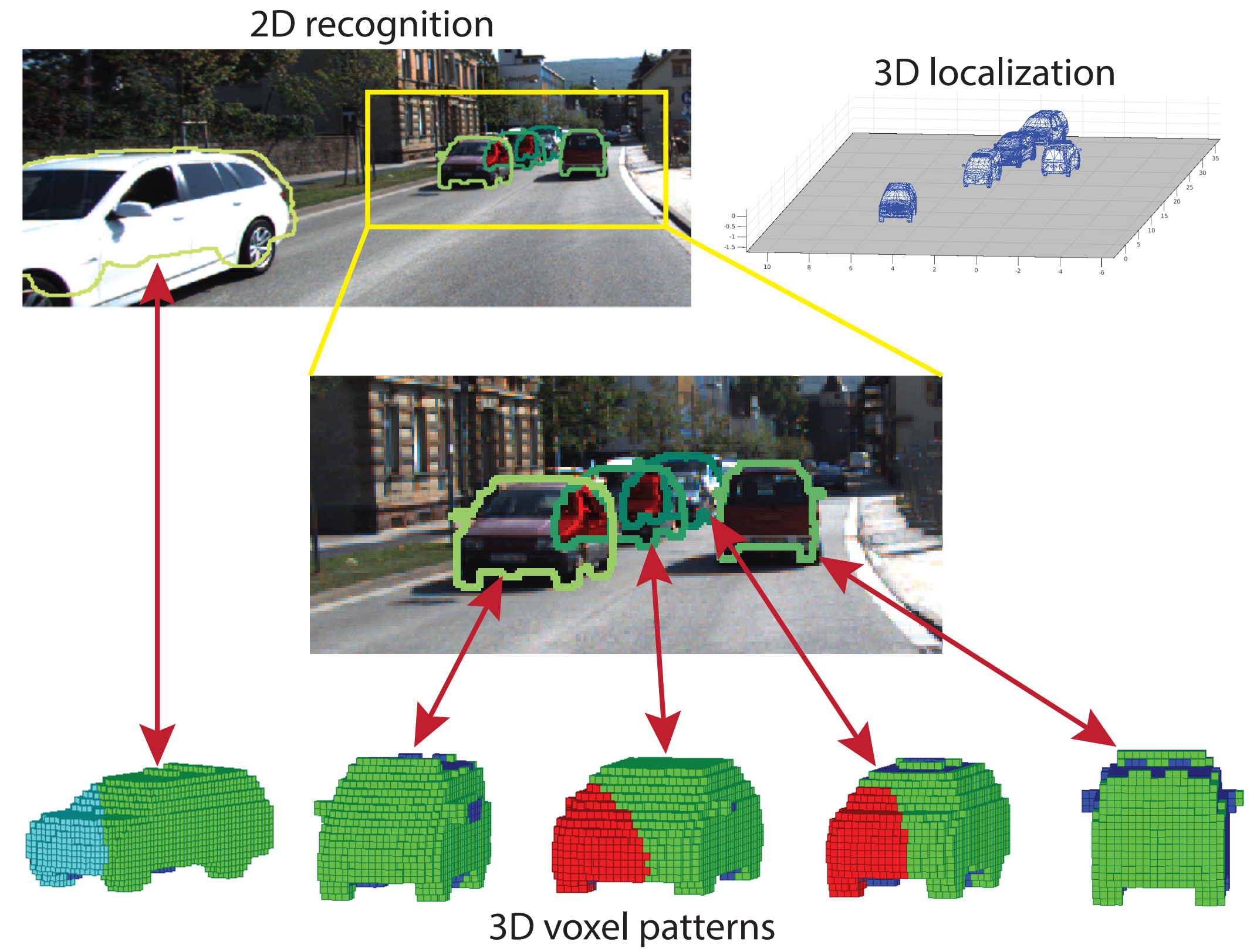}
    \caption{An example of 3D reconstruction for autonomous driving~\cite{xiang2015data}.}
    \label{fig:slam}
\end{figure}

Among existing techniques, formulating the 2D-to-3D projection is an important problem. Depth is thus a crucial piece of information for inferring such a projection. Depth can be obtained from a depth sensor, which can be combined with the RGB camera to form an RGB-D sensor, such as Kinect~\footnote{https://docs.microsoft.com/en-us/windows/apps/design/devices/kinect-for-windows}. 3D reconstruction can be decomposed into two types: 1) 3D reconstruction from multiple views, and 2) 3D reconstruction from a single view.  In recent years, extensive research has been focused on exploring the potential of deep learning to estimate depth in VR/AR, e.g., ~\cite{lai2019real,li2020unsupervised,oh2021bips}.
Depth estimation from the multi-view is a critical task in achieving the Vetaverse. 
The estimated distance directly determines the position of the contents in the immersive environment. The common way to estimate depth is to use a stereo camera~\cite{el2019distance}. 
In VR, stereo depth estimation is conducted in the virtual space. Therefore, depth estimation estimates the absolute distance between a virtual object and the virtual camera (first-person view) or the referred object (third-person view). The traditional methods first extract feature points and then use them to compute the cost volumes for disparity estimation~\cite{scharstein2002taxonomy}.

In XR, one of the critical issues is to ensure that depth estimation is done based on both virtual and real objects. In this way, the XR users can place the virtual objects in the correct positions. 
Early methods in the literature for depth estimation in AR/MR rely on the absolute egocentric depth~\cite{Wildenbeest2013}, indicating how far it is from a virtual object to the viewer. The key techniques include ``blind walking''~\cite{lampton1995distance}, imagined blind walking~\cite{loomis2003visual}, and triangulation by walking~\cite{willemsen2004effects}. Recently, deep learning-based methods have been applied to XR~\cite{prokopetc2019towards,badias2020real}, showing much more precise depth estimation performance. Stereo cameras have been applied to some HMDs, e.g., the Oculus Rift~\cite{kanbara2000stereoscopic}. Infrared camera sensors are also embedded in some devices, such as HoloLens, enabling easier depth information collection.

\subsection{Simultaneous Localization and Mapping (SLAM) }\label{subsec:slam}
Visual SLAM is a pivotal component inside the complex architecture of the autonomous driving system. A visual SLAM algorithm has to solve several challenges simultaneously: (1) unknown space, (2) free-moving or uncontrollable camera, (3) real-time, and (4) robust feature tracking (drifting problem)~\cite{ouerghi2020comparative}. Among the diverse SLAM algorithms, the ORB-SLAM series, e.g., ORB-SLAM-v2~\cite{mur2017orb}, have been shown to work well, e.g., in the AR systems~\cite{zeng2018orb,ouerghi2020comparative}.

Visual SLAM algorithms often rely on three primary steps: (1) feature extraction, (2) mapping the 2D frame to the 3D point cloud, and (3) close loop detection.
The first step for many SLAM algorithms is to find feature points and generate descriptors~\cite{cadena2016past}. Traditional feature tracking methods, such as Scale-invariant feature transform (SIFT)~\cite{lowe2004distinctive}, detect and describe the local features in images; however, they are often too slow to run in real-time. Therefore, most AR systems rely on computationally efficient feature tracking methods, such as feature-based detection~\cite{rublee2011orb} to match features in real-time without using GPU acceleration. Although recently, convolutional neural networks (CNNs) have been applied to visual SLAM and achieved promising performance for autonomous driving with GPUs~\cite{milz2018visual}, it is still challenging to apply to resource-constrained mobile systems.

With the tracked key points (features), the second step for visual SLAM is how to map the 2D camera frames to get 3D coordinates or landmarks, which is closely related to camera pose estimation~\cite{reitmayr2010simultaneous}. The SLAM algorithm first estimates the key points when the camera outputs a new frame. These points are then mapped to the previous frame to estimate the scene's optical flow. As a result, camera motion estimation paves the way for the same key points to be found in the new frame. However, in some cases, the estimated camera pose is not precise enough. Some SLAM algorithms, e.g., ORB-SLAM~\cite{mur2017orb,nerurkar2020system} add extra data to refine the camera pose by finding more key point correspondences. Triangulation of the matching key points from the connected frames generates new map points. This procedure combines the 2D positions of key points in the frames, as well as the translations and rotations between them.  

\begin{figure}[t!]
    \centering
    \includegraphics[width=\columnwidth]{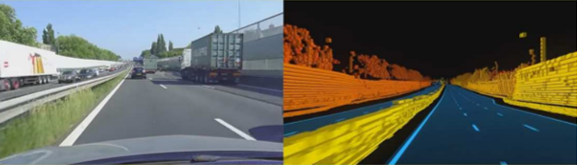}
    \caption{An example of visual SLAM for autonomous driving~\cite{milz2018visual}.}
    \label{fig:slam}
\end{figure}

The last key step of SLAM aims to recover the camera pose and obtain a geometrically consistent map, also called close-loop detection~\cite{biswas2012depth}. As shown in Figure~\ref{fig:slam} for autonomous driving~\cite{milz2018visual}, local map generation or vehicle pose estimation is an essential step for the driving system's scene understanding. ORB-SLAM~\cite{mur2017orb} determines whether key points in a frame match previously detected key points from a different location. If the similarity exceeds a certain threshold, it indicates that the user has returned to a previously visited location. Some SLAM algorithms have recently combined the camera with other sensors, such as the IMU sensor, to improve loop detection precision~\cite{paul2017comparative}, and some works, e.g.,~\cite{schonberger2018semantic}, have attempted to fuse semantic information to SLAM algorithms to ensure loop detection performance. 

Although current SOTA visual SLAM algorithms have established a solid foundation for spatial understanding, Vetaverse requires understanding of more complex environments, particularly the integration of virtual objects and real environments. Hololens has already begun to improve its spatial understanding, and Apple has introduced ARKitv2\footnote{\url{https://developer.apple.com/videos/play/wwdc2018/602}} for 3D keypoint tracking. In Vetaverse, the perceived virtual universe is built in the shared 3D virtual space. As a result, acquiring the 3D structure of an unknown environment and sensing its motion is critical but difficult. This could aid in data collection for purposes such as digital twin construction, which can be linked with AI to achieve auto conversion from the physical world. Furthermore, it is critical in Vetaverse to ensure the accuracy of object registration and interaction with the physical world. With these stringent requirements, we anticipate that Vetaverse's SLAM algorithms will become more precise and computationally efficient.

\subsection{XR}
XR is a shared virtual space that combines physical reality, AR and VR. This allows users to work, play, and interact socially in the blended post-reality universe, which is a web of social, networked immersive environments on persistent multi-user platforms. 

AR is a set of technologies that superimpose digital items (e.g., visual components, sound, and 3D objects) on the physical world. It spatially augments the physical layer with virtual objects using mobile devices such as smartphones, tablets, glasses, or other transparent surfaces such as the vehicle's windshield~\cite{mystakidis2022Metaverse}. AR bridges the gap between all the digital information we have and the real world where we use it.  It extends the user's capability to take advantage of the torrent of information and insights produced by billions of smart, connected devices (i.e., IoTs). AR is becoming the new interface between humans and vehicles, bridging the digital and physical worlds. 

AR-assisted ADAS~\cite{wang2020augmented} provides the drivers with visual guidance using information computed cooperatively by multiple connected vehicles. This system uses AR as a human-machine interface (HMI), which puts guidance information on top of what the driver can see through the windshield. This becomes an effective solution in intersections that are not served by traffic signals. In these cases, the driver can cross the intersection under the AR guidance without any full stop. This reduces travel time and energy consumption in a safer, more efficient, and more comfortable way. EAVVE~\cite{9163287} and AICP~\cite{aicp} push AR-assisted ADAS forward towards reality via edge-offloading provision and lightweight information filtering, to compensate for the less-powered vehicles and remove noisy visual data. 

Virtual reality (VR) creates separate digital and artificial environments in which users feel immersed as if they are in a different world, and behave as they would in real life. Using head-mounted displays (e.g., Oculus Quest 2), VR platforms (e.g., Mozilla Hub\footnote{\url{https://github.com/mozilla/hubs}}, Spatial\footnote{\url{https://spatial.io/}} and Meta Workrooms\footnote{\url{https://www.oculus.com/workrooms}}) extend the user's field of view (FOV), block the entire ambient, and offer an immersive virtual environment independent of the user’s actual surroundings. These environments are completely virtual and intensify fictional realities. They are based on 3D models of the virtual space and objects in which 3D models of vehicles or pedestrians (i.e., avatars) navigate.

VR has been utilized to demonstrate the safety of L3+ automated driving with respect to human drivers. Owing to the need of test driving up to several million miles, simulations and high-fidelity virtual environments are employed to efficiently achieve the testing of autonomous vehicles. Such testing requires two primary components in a synthetic environment, which are validated sensor models and noise models for each sensor technology. The sensors feed the vehicle’s information into the automated system to plan its trajectory and navigate safely. Espineira et al~\cite{espineira2021realistic} propose innovative real-time LiDAR-based and probabilistic rain models. The model is based on the Unreal engine\footnote{\url{www.unrealengine.com/}} and runs in real-time, synchronized with the visual rendering, in an immersive driving simulator.

\subsection{Network}
Dedicated Short Range Communication (\textbf{DSRC}) is a vehicular communication technology  for vehicular communication that typically operates in licensed spectrum in the 5.9 GHz band in several countries including the united states~\cite{kenney2011dedicated}. \textbf{ITS-G5} is an analogous European technology that is developed by the European Telecommunications Standards Institute (ETSI) to ensure interoperability among communication devices from different manufacturers~\cite{ETSI.EN.302.663.V1.2.0}. DSRC and ITS-G5 allow vehicles and RSUs to form vehicular ad-hoc networks (VANETs) through vehicle-to-vehicle (V2V) and infrastructure-to-vehicle (V2I) communications. They also communicate with road infrastructure (V2I), pedestrians (V2P), and data networks (V2N)~\cite{wang2017overview}. The bandwidth offered by DSRC and ITS-G5 is limited, 27 mb/s, which does not meet the minimum bandwidth demand of Vetaverse applications.

3GPP has developed LTE Cellular-V2X (\textbf{LTE C-V2X}) to operate in 5.9 GHz band (similar to DSRC) in addition to the licensed carriers via network infrastructure. This enables direct communications in the absence of cellular infrastructure in a distributed manner~\cite{hakeem20205g}. LTE C-V2X operates in two transmission modes: 1) C-V2X/PC5 which supports V2X direct sidelink communications, allowing vehicles and RSUs to inter-communicate directly without the need for infrastructure, and thus providing lower delay, higher throughput, and lower energy consumption and better spectral efficiency~\cite{chen2017vehicle}, 2) C-V2X/Uu communications to connect road users (e.g., vehicles and RSUs) indirectly through LTE infrastructure. In this mode, since the V2X transmissions are scheduled, interference and collisions are lessened~\cite{hakeem20205g}. Recently, 3GPP has updated C-V2X to leverage the fifth-generation (5G) mobile communications standard, thus leading to New Radio (NR) C-V2X which is compatible with the evolution of 5G. \textbf{NR C-V2X}  provides ultra-reliable low-latency communications and ultra-high throughput (10-100$\times$ higher compared to LTE).  Similar to LTE C-V2X, NR C-V2X enables direct communication between vehicles, and indirect communications (via infrastructure). Furthermore, 5G supports the integration of non-3GPP communication and telecommunication systems including WiFi variants, ZigBee, and Bluetooth, bringing more flexibility to vehicular networks. For instance, it allows vehicles, drivers, passengers, and pedestrians to leverage the most suitable system for their selected application~\cite{shah20185g}.

While mobile edge computing (MEC) can enable responsive Vetaverse services, high-capacity mobile edge networking is indispensable to meet the bandwidth and latency demand. However, the fourth (4G) and fifth generations (5G), which are based on sub-6 frequency bands, fall short to satisfy the bandwidth demand for these services. Vetaverse applications such as VR 360° video streaming are bandwidth-hungry and latency-sensitive high-quality video streaming

The \textbf{sixth generation (6G)} promises to combine ground-breaking technologies to attain a capacity of at least 1 Tb/s and peak data rates of 10 Tb/s.  6G aims to combine multiple frequency bands in the electromagnetic spectrum to meet the Vetaverse demand. These bands are primarily radio frequencies of millimeter waves (mmWave) and Terahertz (THz). Besides, 6G is expected to utilize non-orthogonal multiple access (NOMA) by applying different levels of power. Using NOMA, 6G communication systems can increase the bandwidth capacity and enhance spectral efficiency, thus providing high data rates and ubiquitous connectivity. As mmWave and THz-based communications require a line of sight (LoS) between the Vetaverse user and the access point, intelligent reflecting surfaces (IRS) can tune the communication beams towards the user when obstacles block the LoS, thus enhancing communication reliability~\cite{yu20226g}.

\subsection{Edge and Cloud}
Vetaverse applications are computation intensive and present challenges for onboard capacities. Offloading thus is a good method to ensure speedy processing and a good user experience. Traditional cloud, although can provide long-term large-scale data maintenance, is hindered by latency and network congestion to provide offloading services for real-time user interaction. Edge computing computes, stores, and transports data closer to end-users and their devices, reducing latency. Satyanarayana et al.~\cite{cloudlet} observed in 2009 that providing cloud-like infrastructure just one wireless hop away from mobile devices may revolutionize the game. As such, edge can provide for real-time application offloading such as object detection for ARHUD~\cite{9163287} and cloud can provide large-scale orchestration like traffic monitoring and scheduling.

\subsection{HD Map}
Map plays a crucial role in automatic driving. It compensates for the circumstances when the sensor is blocked or the range is insufficient. In the meantime, it provides the correct interpretation corresponding to the understanding of the scene. Furthermore, knowledge of the map can be inherited from previous driving tracks. With the continuous progress in the field of automatic driving, the maps are of necessity to be more and more precise, while fulfilling higher quality requirements. Although rough road routes are 
adequate for navigation equipment, maps must provide more diversified information for automatic driving. Therefore, the high-definition (HD) map, which is dedicated to automatic driving, is an indispensable component in all major automated driving projects~\cite{MakingBD}. Unlike traditional navigation maps, HD maps not only provide road-level navigation information but also provide lane-level navigation information. It is far superior to the traditional navigation map in terms of information richness or accuracy~\cite{Lanelet2AH}.


\begin{figure}[t!]
    \centering
    \includegraphics[width=0.9\columnwidth]{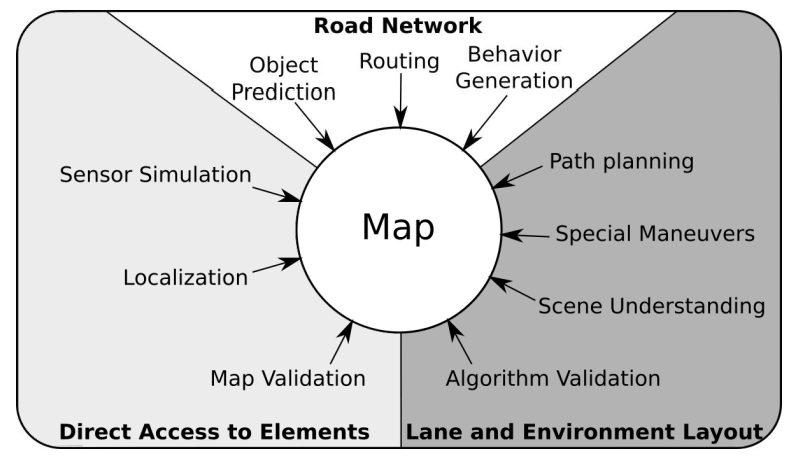}
    \caption{This figure shows the application of HD map.}
    \label{fig:HD map app}
\end{figure}

\subsubsection{Point Cloud Map} 
\begin{figure}[t!]
    \centering
    \includegraphics[width=0.9\columnwidth]{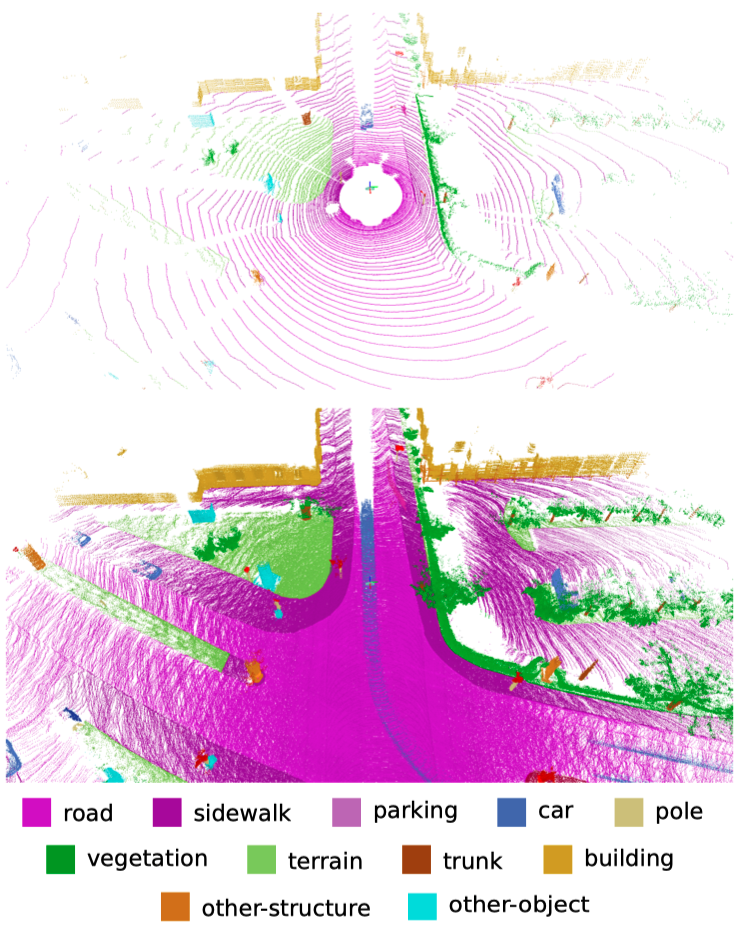}
    \caption{A sample of point cloud map from \cite{behley2019semantickitti}}
    \label{fig:05-pcm-1-example}
\end{figure}
One of the most popular categories of HD map is point cloud maps. 
A point cloud is a set of data points in space. 
These points can represent 3D shapes or objects.
Each point location has its own coordinate set $(X, Y, Z)$, and some point clouds also have color values $(R, G, B)$ and reflection intensity values~\cite{wan2021rgb,behley2019semantickitti}.
Point clouds are usually generated by 3D scanners, which measure many points on the surface of surrounding objects. 
Self-driving vehicles have many LiDARs that can obtain a large number of point clouds outside the vehicle, as shown in Figure~\ref{fig:05-pcm-1-example}.
These point clouds can be used for visualization, rendering, and semantic understanding of the environment outside the vehicle.
The maps formed by these point clouds have important applications in the Vetaverse.

We believe that the application of point cloud maps in the Vetaverse has three points. 
1. Outside reconstruction: point cloud maps can provide data for 3D reconstruction of the outside environment. 
2. The registration of point cloud data of adjacent vehicles provides a potential possibility for vehicle-to-vehicle communication. 
3. The collection of point clouds from multiple vehicles lays the foundation for building a global map of the educational system.

For 3D reconstruction, many techniques on point clouds have application value. 
These 3D reconstruction techniques based on point clouds can be roughly summarized into the following four processes. 
1. Data preprocessing: such as pose correction, coordinate unification. 
The purpose of this step is to remove some problems caused by the hardware, such as inconsistent coordinate systems, noise, etc.
2. Filtering~\cite{zhang2020pointfilter}, denoising~\cite{luo2021score,zeng20193d}, registration~\cite{pomerleau2015review,yang2020teaser}, segmentation~\cite{nguyen20133d,landrieu2018large} of point cloud data. The purpose of this step is to segment the point cloud into different parts, e.g., distinguishing the point clouds of street trees, vehicles, people, and roads.
3. Gridding of point cloud, this step will complete the conversion of point cloud to mesh~\cite{xie2020grnet}. 
4. Panoramic texture map, which maps information such as color to the mesh.

The 3D reconstructed map of the point cloud can map the real-world scene into the Vetaverse, and prompt the driver according to the semantic information of the scene. 
For a single vehicle, the driver can get prompts from the Vetaverse system in real time, e.g., obstacles detected in point cloud maps.
In the case of multiple vehicles, different vehicles can broadcast their own driving routes through the shared point cloud map.
This is extremely valuable for some attack scenarios, such as ambulances and police cars in emergency scenarios.
For the global traffic system, the point cloud maps obtained by all vehicles are used to form a global map, which can provide important information in traffic planning and route setting.

\subsection{Digital Twin}

As an integral part of a city, the performance of the transportation system is inseparable from city activity. Therefore, a more intelligent transportation system~(ITS) is essential for urban development. With the gradual maturation of digital twin technology, the construction of digital twins of ITS has become essential to provide efficient transportation operation, analysis, presentation, simulation, prediction, and planning~\cite{liu2022vision}. The digital twin of vehicles, pedestrians, roads, and traffic signals allows for the visualization of real-time dynamics, and 3D models enable the observation of space from multiple angles~\cite{chen2018digital}. Transferring transportation demand and traffic flow into the management system can effectively reduce accidents and violations, improve transportation efficiency, and improve residential well-being~\cite{RUDSKOY2021927}~\cite{creb2022intelligent}. Furthermore, the continuously collected data increases the degree of similarity between the digital twin and the real world, enabling the development of a twin-based simulation environment for autonomous driving. 3D model reconstruction can generate synthetic data similar to realistic camera, LiDAR and other sensors, based on learning from historical traces, thus drastically reducing the cost of data collection for developing autonomous driving algorithms~\cite{airsim2017fsr,Dosovitskiy17}. Leveraging advanced communication networks and sensory-and-image data fusion techniques, bidirectional real-time data exchange between physical entities and digital twins can achieve seamless bidirectional communication and mutual control, i.e., realizing the digital shadow of ITS.

\begin{figure*}[t!]
    \centering
    \includegraphics[width=0.45\textwidth]{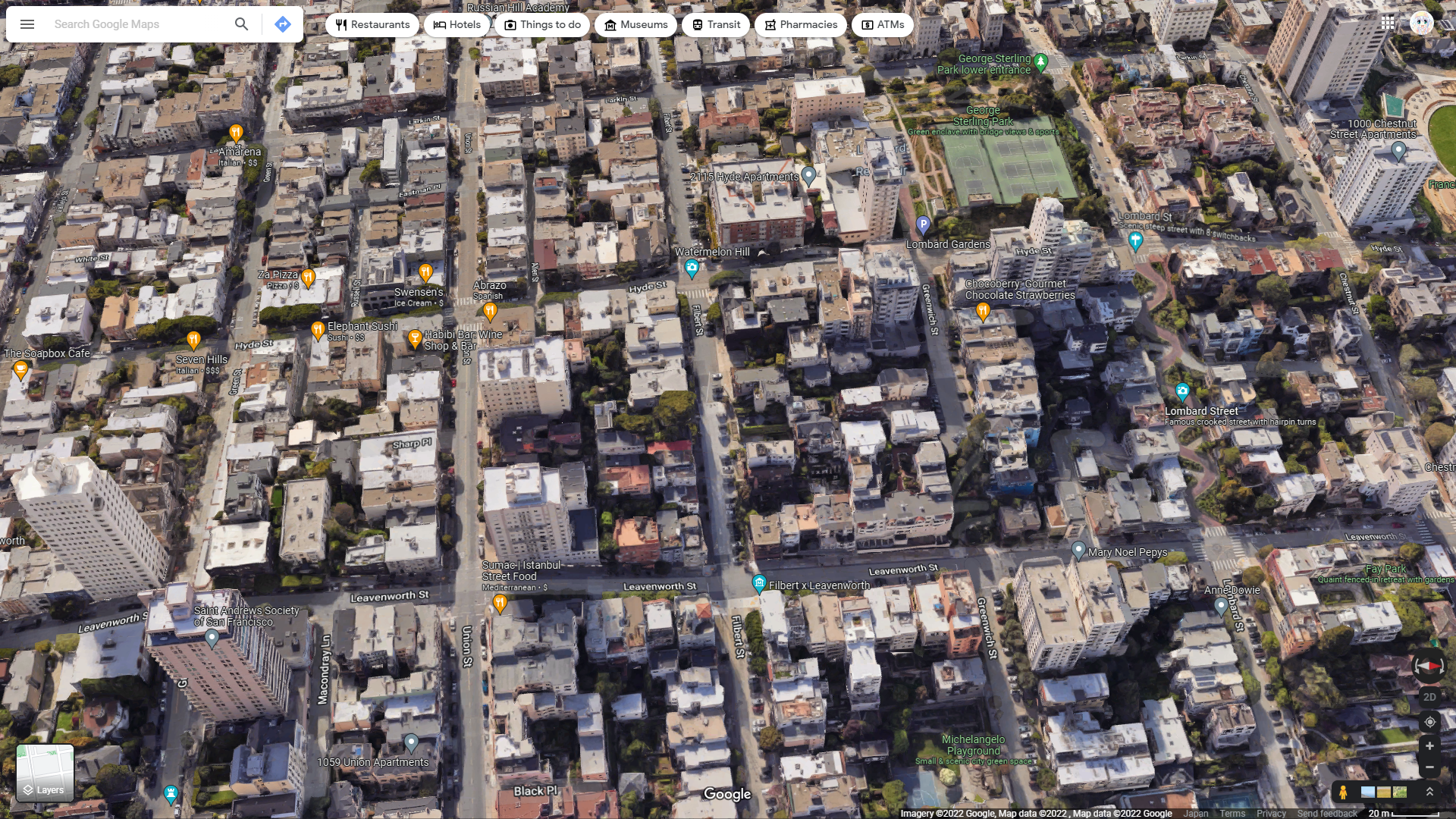}
    \hspace{0.25em}
    \includegraphics[width=0.45\textwidth]{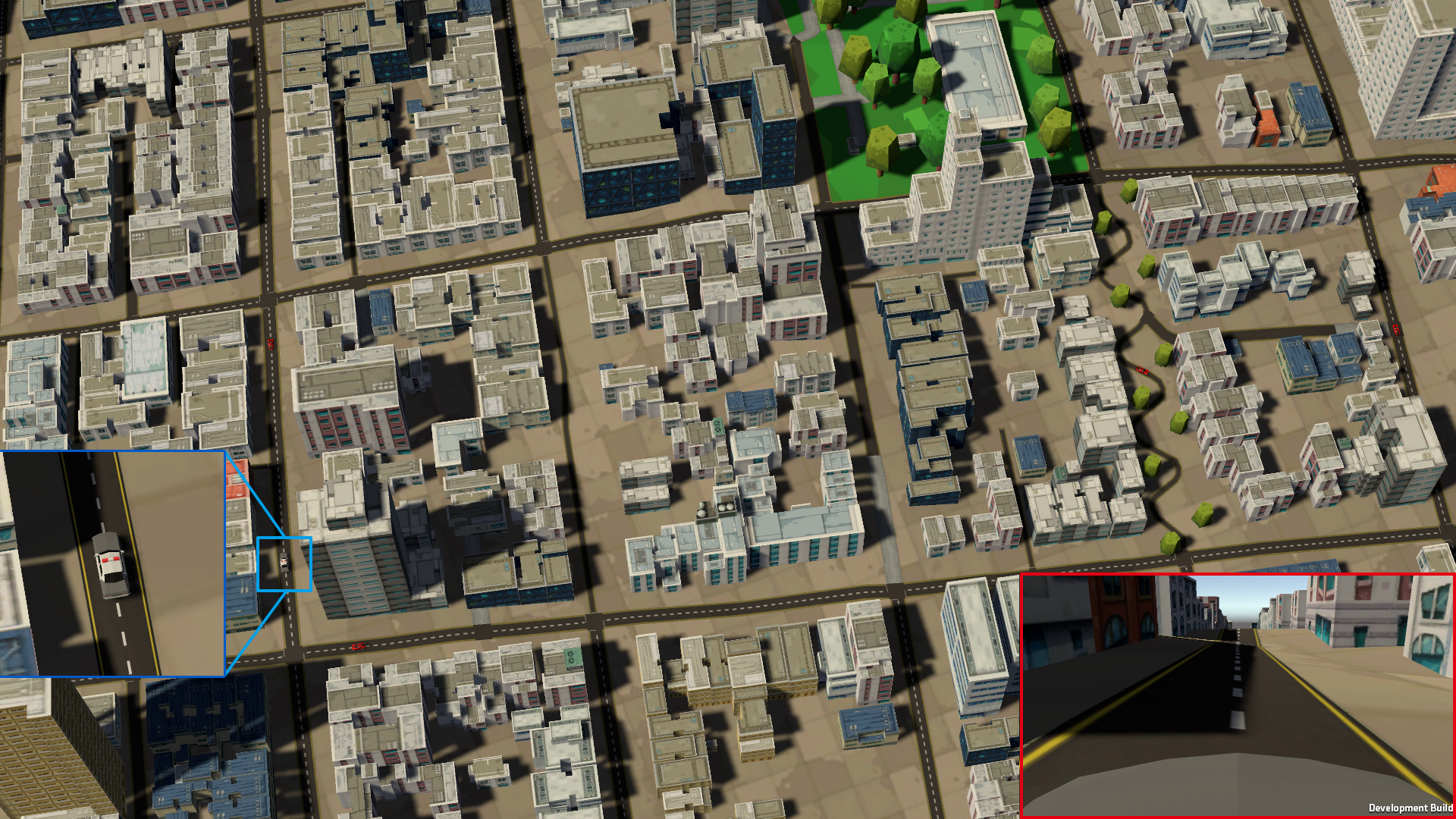}
    \caption{An example of a 3D digital twin of ITS (the image on the right side), showing vehicles traveling in a city and a first-person view of one car (the bottom right rectangle). The image on the left is a screenshot of real world map (San Francisco, Lombard Street, from Google Map).}
    \label{fig:CarInCity}
\end{figure*}


\section{TS-Metaverse}\label{sec:its}
In this section, we present the vision of TS-Metaverse and discuss core aspects in detail, including the digital twin of the system, XR for connected vehicles, and resource orchestrations. 

\subsection{Intelligent Transportation Systems (ITS) and Metaverse}
\label{subsec:ITS}
\textbf{Intelligent Transport Systems (ITS)} are novel transport systems that leverage information, communication, and control technologies (ICT) to link humans, vehicles, and roads. ITS aims to resolve a variety of road traffic issues by enabling smart, efficient, and safe transportation~\cite{makino2016intelligent}. As part of enabling ICT, Automotive IoT is the primary component that has become a prominent hotspot for diversified multi-purpose applications. For instance, in-vehicle IoT assists drivers with driving, braking, parking, and lane-changing activities. It also takes on-the-spot decisions while partly controlling the vehicle operations to avoid accidents and reduce the load on the driver. In-vehicle IoTs integrate multiple sensors such as proximity sensors, LiDAR, radar, and cameras with IoT systems to reduce human error and make driving safer and more convenient. The pervasive deployment of communication technologies such as cellular vehicle-to-everything (C-V2X)~\cite{hakeem20205g}, dedicated short-range communication (DSRC)~\cite{kenney2011dedicated}, and the European standard (ITS-G5)~\cite{ETSI.EN.302.663.V1.2.0}, has enabled networking among the vehicles and with the infrastructure. In particular, vehicles intercommunicate, vehicle-to-vehicle (V2V), in an ad-hoc manner (VANET). They also communicate with road infrastructure (V2I), pedestrians (V2P), and data networks (V2N)~\cite{wang2017overview}.  

\textbf{TS-Metaverse.} With the advent of the Metaverse, physical and cyber worlds can be blended into mixed physical-virtual environments. The real-world data collected by IoT devices and sensors installed in vehicles, on pedestrians’ mobile phones, and along the roads is the key enabler for synchronizing the two worlds. For example, location data from GPS sensors and vehicle characteristics can aid in transforming them into digital twins in the Vetaverse. Accordingly, vehicles can be represented as dynamic objects, while buildings, roads, and road infrastructure are represented as stationary objects in the virtual space. Humans (e.g., drivers, passengers, and pedestrians) can use avatars to represent themselves and communicate with each other and the virtual environment. Data freshness is crucial to maintaining a well-synchronized digital twin in the Vetaverse, i.e., acquiring the most up-to-date data from the road users (e.g., vehicles, pedestrians) and road-side sensors (e.g., inductive loop detectors, traffic lights) in the physical world to keep the digital twin updated in real-time~\cite{han2022dynamic}.

\subsection{Digital Twin} 
\begin{figure*}[t!]
    \centering
    \includegraphics[width=0.95\textwidth]{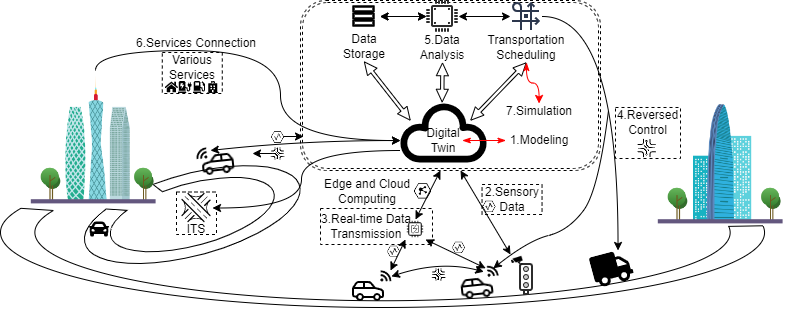}
    \caption{An example of a digital twin of an urban transportation system, which can transcend some realistic constraints and provide full support in the utilization of computing resources, route planning, connection of users to services, statistical analysis of data, etc.}
    \label{fig:VDT}
\end{figure*}

Currently, digital twin technology is widely used in industry, particularly for real-time automated monitoring, control, and management, and there is a high demand for these applications in increasingly complex urban and rural transportation systems. In the TS-Metaverse, the construction of a high-precision digital twin can serve as the foundation for high-precision mapping and navigation services, accident analysis and prediction, and autonomous driving simulations, which can accelerate the development of ITS~\cite{RUDSKOY2021927}. An example of a digital twin of an urban transportation system is shown in Figure~\ref{fig:VDT}.

\begin{figure*}[t!]
    \centering
    \includegraphics[width=0.95\textwidth]{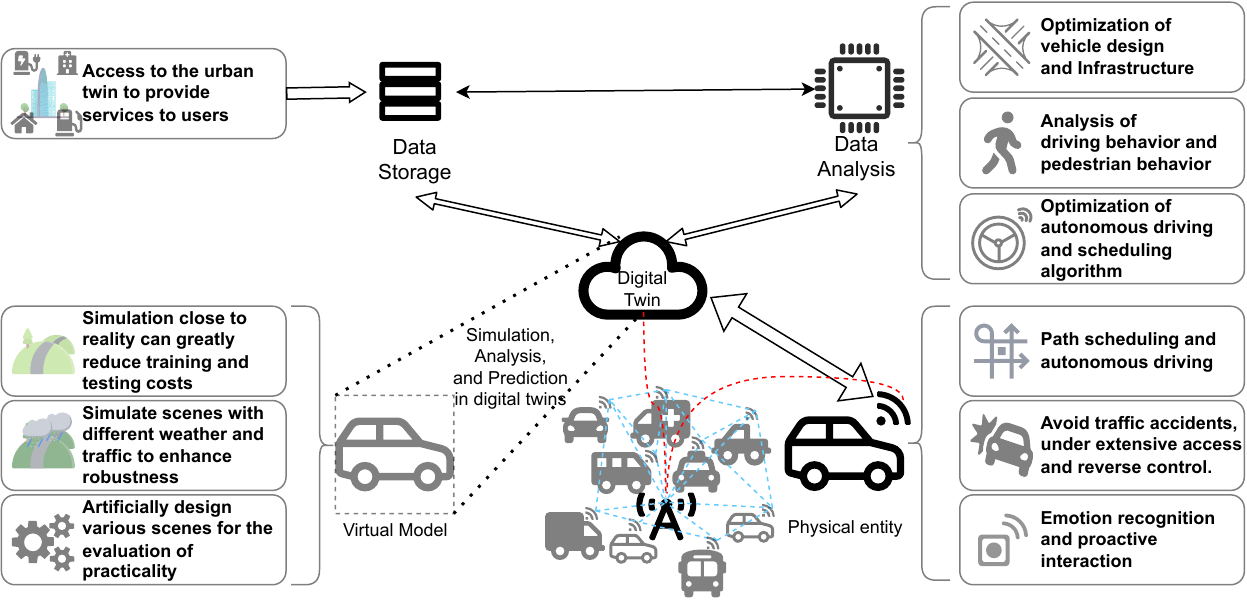}
    \caption{After establishing the digital twin, stable and effective algorithms can be used to reversely control the controllable physical entity by driving the virtual twin in real time. Thus, user experience and driving safety can be improved based on large-scale data analysis and wide-scale information collection.}
    \label{fig:ReversedControl}
\end{figure*}

As Figure~\ref{fig:VDT} depicts, a digital twin of ITS can be constructed according to the following phases:
\begin{enumerate}[leftmargin=*]
\item  \textbf{Modeling} is the first step in creating a digital twin. Traditional traffic simulation systems have a tendency to oversimplify the environment and traffic objects, making it challenging to achieve comprehensive utilization of the collected data. Roads, traffic signals, traffic signs and markings, vehicles, road users, transportation-related services, the natural environment, and weather conditions are the ITS objects to be modeled. We need to construct their geometric models, physical models, behavioral models, and rule models to accurately reflect their actual conditions in the virtual world~\cite{tao2019five}. Technologies such as 3D scanning, 3D reconstruction, and geographic mapping are essential for geometric modeling. Rule modeling necessitates the collection of extensive data. Meanwhile, detailed physical modeling and behavioral modeling require more manual intervention, which remains an open challenge~\cite{8972429}.


\item \textbf{Sensor data acquisition.} Using GPS, LiDAR, cameras, traffic signals, and other sensor devices to collect various types of data necessitates collecting and filtering traffic information comprehensively and effectively through edge computing, IoT and other technologies, and securely transferring the useful information to the digital twin platform~\cite{tao2018digital,s18041212}.

\item \textbf{Real-time data transmission and model updates.} The collected data is transferred to the platform via advanced communication networks in real-time to update the digital twin. Data processing faces challenges such as the harmonization of diverse protocols and standards, data fusion, information prioritization, and noise pruning~\cite{XIA2021210} etc.

\item \textbf{Reversed control.} Figure~\ref{fig:ReversedControl} shows the application and advantages of reverse control. The digital twin should be able to control the physical entity (at least to some extent)~\cite{JIANG202136}, which is a key difference from traditional virtual model solutions. ITS can guide real traffic by directing the digital twin, for example, to improve driving safety. Better, in a digital world, we can predict the impact of each object's behavior in advance.


\item \textbf{Data analysis.} The digital twin can not only reflect the current state of the physical entity but also store and replay historical traces, which is extremely useful for data analysis. The traffic information obtained in real-time can be used to optimize the deployment of traffic infrastructures, and the characteristics of vehicle and pedestrian movements can be used for simulation and prediction.

\item \textbf{Services connection.} The digital twin of ITS can be linked with various related services such as parking lots, restaurants, shopping malls, etc., to effectively improve user convenience during the commute.

\item \textbf{Simulation} is an important function of the ITS digital twin, which can provide great help in the training and testing of autonomous vehicles. The digital twin combines sensory data with map information to reconstruct realistic scenarios in a digital platform, which is capable of performing high-precision simulations of various roads, traffic, light, weather, and other conditions. This can greatly enhance the generalizability of self-driving algorithms within faster development cycles while reducing training and testing costs.
\end{enumerate}

\subsection{XR for Connected Vehicles}
Lots of jams are caused by accidents and greatly impact the efficiency of ITS. A major cause of accidents is blind spots caused by view obstruction and unfavorable weather conditions. Extended vision can efficiently reduce blind spots and thus is of great importance for accident avoidance, driving safety, and transportation system efficiency. Currently, a suite of technologies can be utilized to actualize the extended vision for automobiles and drivers. Following, we discuss how to utilize XR for connected vehicles to expand the scope of vision and context comprehension.

Most automobiles nowadays are outfitted with sensors and display devices that provide the driver with helpful information, such as about the driving environment and driving habits. The most common form of display is a heads-up display (HUD). Recently, academic and commercial interest in combining AR and HUD has been increasing. AR enables the incorporation of 3-D data images into the HUD rendering background, enabling precise obstacle detection and emergency notifications. Relevant research has focused on various problems, such as cognitive utility, visibility, and the alignment of embedded information with the actual environment \cite{park2013efficient,park2013vehicle}. Connecting vehicular perspectives via ARHUD is in its infancy, though.
Indeed, it is already a challenging matter when only two vehicles share the same vision. Extending this model to a large number of vehicles on the road in real-time would significantly increase the system's complexity. Numerous ambient broadcast messages are transmitted to autos in this scenario. Vehicles must select just the necessary data to prevent information overload and adverse consequences (driver distraction, performance declines, network congestion).

\begin{figure}[t!]
    \centering
    \includegraphics[width=0.7\columnwidth]{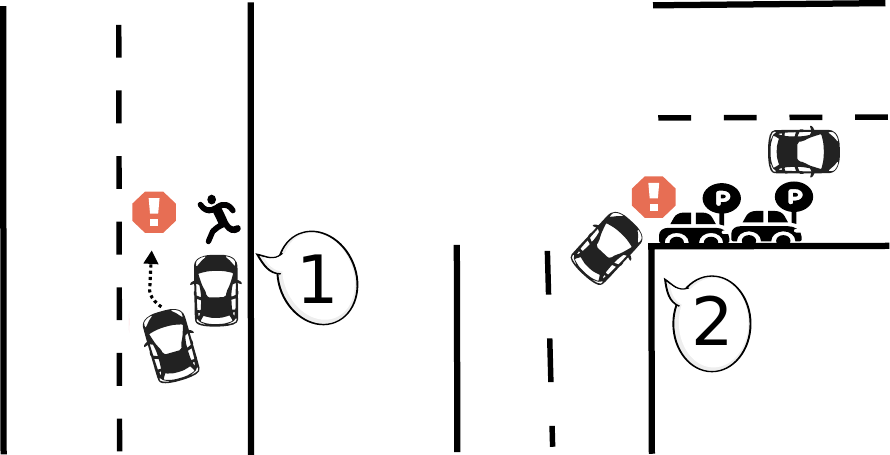}
    \caption{Two accidents that can be avoided with connected vehicular vision~\cite{9163287}.}
    \label{fig:accident}
\end{figure}

But the effort is definitely worthy. For example, Figure~\ref{fig:accident} illustrates two different car incidents that the poor visibility contributed to: (i) The leading vehicle's driver suddenly notices a pedestrian crossing the street quickly and thus hits the brake. However, the following vehicle's driver, whose vision is obstructed by the leading vehicle, is unaware of the pedestrian and is just about to pass the leading vehicle, which is quite a common case and often causes a tragic collision. (ii) Out of the line of sight of the vehicles turning right, some vehicles are parked just around the corner. The leading vehicle luckily took a big curve and avoided the collision. However, the following vehicle is taking a small curve and it is too late to brake when the driver finds out about the parked vehicles~\cite{arve}. 

In the past decade, there have been some efforts focused on extending vehicular vision, utilizing terms such as ``see-through'', ``cooperative vision'', and ``cooperative perception'', among others. Kim et al. proposed a paradigm for general vehicular cooperative perception that tackles several important challenges in the field, such as map merging, communication uncertainty, and sensor multi-modality \cite{kim2013cooperative}. In addition, to enable drivers to share local sensory data for collective perception, \cite{gunther2016collective} proposed and analyzed several communication formats based on ETSI ITS 5G \cite{etsi-its-g5-5g}. After that, they developed an engineering-feasible multimodal cooperative perception system \cite{kim2014multivehicle} and broadened their study with a mirror neuron-inspired algorithm for intention awareness in cooperative autonomous driving \cite{kim2016cooperative}. Garlichs et al. \cite{garlichs2019generation} presented a set of generating rules in 2019 to reduce the transmission burden while retaining perception capabilities. This idea was subsequently incorporated into the ETSI standard \cite{etsi-tr-103-562}. Thandavarayan et al. conducted an in-depth examination of the message-generating rules after reviewing the ETSI standards \cite{thandavarayan2019analysis}. Under the current standards, they assessed the trade-off between communication performance and perception skills and determined that additional optimization is required to reduce information redundancy. To achieve this, Aoki et al. \cite{aoki2020} employed an approach of deep reinforcement learning, selectively transmitting information about objects that were unlikely to have been directly observed by neighboring vehicles.
Some related works have been done with ARHUD. For instance, \cite{qiu2018avr} investigates the sharing of augmented vision between two vehicles. Due to stringent outdoor constraints such as bandwidth, latency, and computing resources, scalability was not addressed. However, the effort did provide answers for fundamental view transformation. EAVVE \cite{9163287} is a comprehensive V2X framework for the instantaneous sharing of enhanced contextual information to allow real-time emergency detection and notification by utilizing the low latency of edge servers. 

\begin{figure}[t!]
    \centering
    \includegraphics[width=0.7\columnwidth]{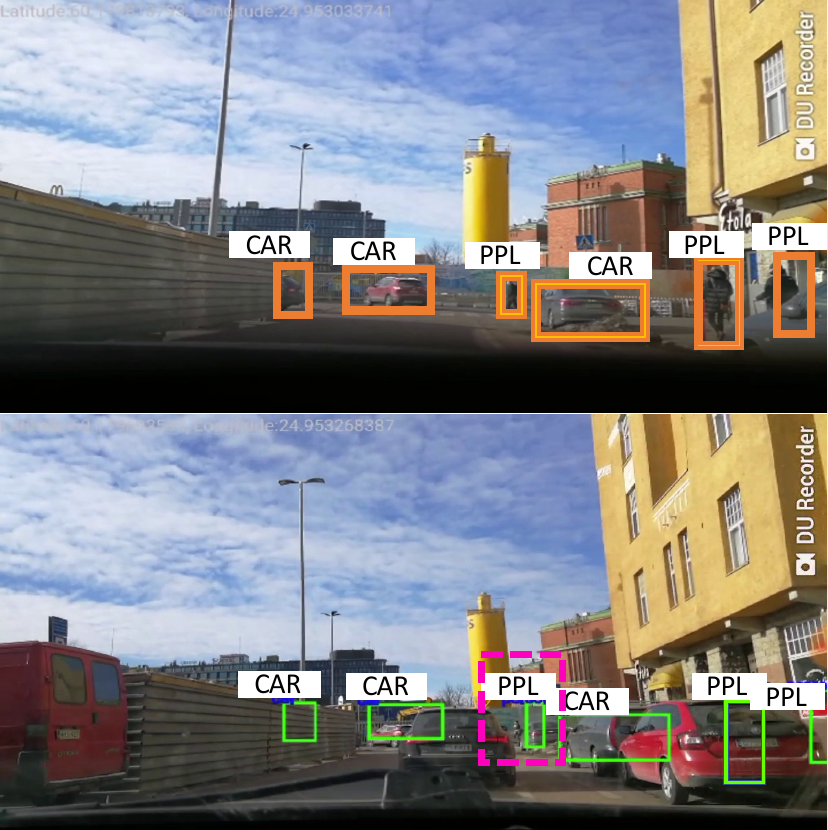}
    \caption{A depiction of a simplistic cooperative perception system. A vehicle in the lead identifies things indicated by orange bounding boxes (top figure). These objects are displayed in green bounding boxes to the driver of the following vehicle (bottom figure). AICP filters these items to only display critical ones (such as the pedestrians in the pink bounding box) in order to limit the number of objects displayed on-screen. }
    \label{fig:filter}
\end{figure}

Connected vehicles face numerous challenges, including timely and synchronized information distribution, data fusion, communication overhead, and an often-overlooked one: the shared information may be overwhelming for both the driver and the vehicle's decision-making system, resulting in driver distraction rather than assistance.
Figure \ref{fig:filter} depicts an example of a following vehicle's vision in a leading-following vehicle situation utilizing naive cooperative perception and augmented reality (AR). The vehicle in the lead detects objects within its field of view and relays information about them. Based on received signals, the following vehicle calculates position transformations and renders all green-boxed objects. When such a system is extended to city-block-level perception without information filtering, drivers are faced with an abundance of items. This hinders their vision, and consequently, their driving experience is diminished. In reality, the time it takes a driver to make a decision increases logarithmically as the number of stimuli or objects increases \cite{liu2020relevant}. In addition, it is crucial to limit the number of objects to the cognitive capacity of humans, which is approximately 7$\pm$2 items \cite{Schweickert1986ShorttermMC}. Therefore, cooperative perception requires efficient filtering. Such a system may, for instance, recognize the things indicated in the pink box in Figure \ref{fig:filter} as essential and display their data while ignoring the remaining items. Augmented Informative Cooperative Perception (AICP) presents the first answer to this challenge. AICP is a fast-filtering solution that maximizes the informativeness of shared visionary data between vehicles to improve the fused presentation at the cost of an appropriate amount of additional delay~\cite{aicp}.

\subsection{Resource Orchestration with Edge and Cloud}
TS-Metaverse can greatly improve the efficiency of the transportation system and benefit the drivers by enhancing visionary and context-understanding capabilities. However, the bulk of those tasks are computationally intensive and battery-consuming. To alleviate these burdens, offloading is a pivotal step, particularly for the ordinary vehicles that dominate the roadways. But it comes at the expense of increased network latency. A balanced trade-off is essential in order to make the offloading process transparent to the user experience. For instance, to better understand the surroundings, a vehicle needs to process captured images (or point clouds, etc.) at a higher frequency to get more granular information, for which offloading results in longer latency. 

Due to the fluctuating and unpredictable high latency \cite{cloudlet,9163287}, cloud offloading cannot consistently achieve the ideal balance and results in long-tail latency performance, which harms the user experience \cite{40801}. Recent cloud reachability measurements indicate that the present cloud distribution can provide network latency of less than 100 milliseconds, meeting ordinary vehicular application requirements. However, in wireless networks, only China (out of 184) satisfies the delay requirement for more delay-sensitive in-car applications such as AR, VR, and emergency warning. Therefore, a complementary solution is required to ensure a smooth and safe driving experience.

Edge computing, which computes, stores, and transmits data closer to end-users and their devices, can minimize user-experienced latency relative to cloud offloading. Satyanarayanan et al. \cite{cloudlet} identified in 2009 that installing cloud-like infrastructures just one wireless hop away from mobile devices, i.e., so-called cloudlets, may revolutionize the game, as demonstrated by several subsequent efforts. Specifically, Chenemph et al. \cite{10.1145/3132211.3134458} examined the latency performance of edge computing by conducting practical tests on various applications. They demonstrated that LTE cloudlets might deliver considerable benefits (60 percent reduced latency) over cloud offloading by default. Similarly, Haemph et al. \cite{ha2014towards} discovered through measurements that edge computing may cut service latency by at least 80 ms on average compared to the cloud, which is crucial for real-time in-car applications. 

\begin{figure}[t!]
    \centering
    \begin{subfigure}[b]{0.5\textwidth}
         \centering
         \includegraphics[width=.8\columnwidth]{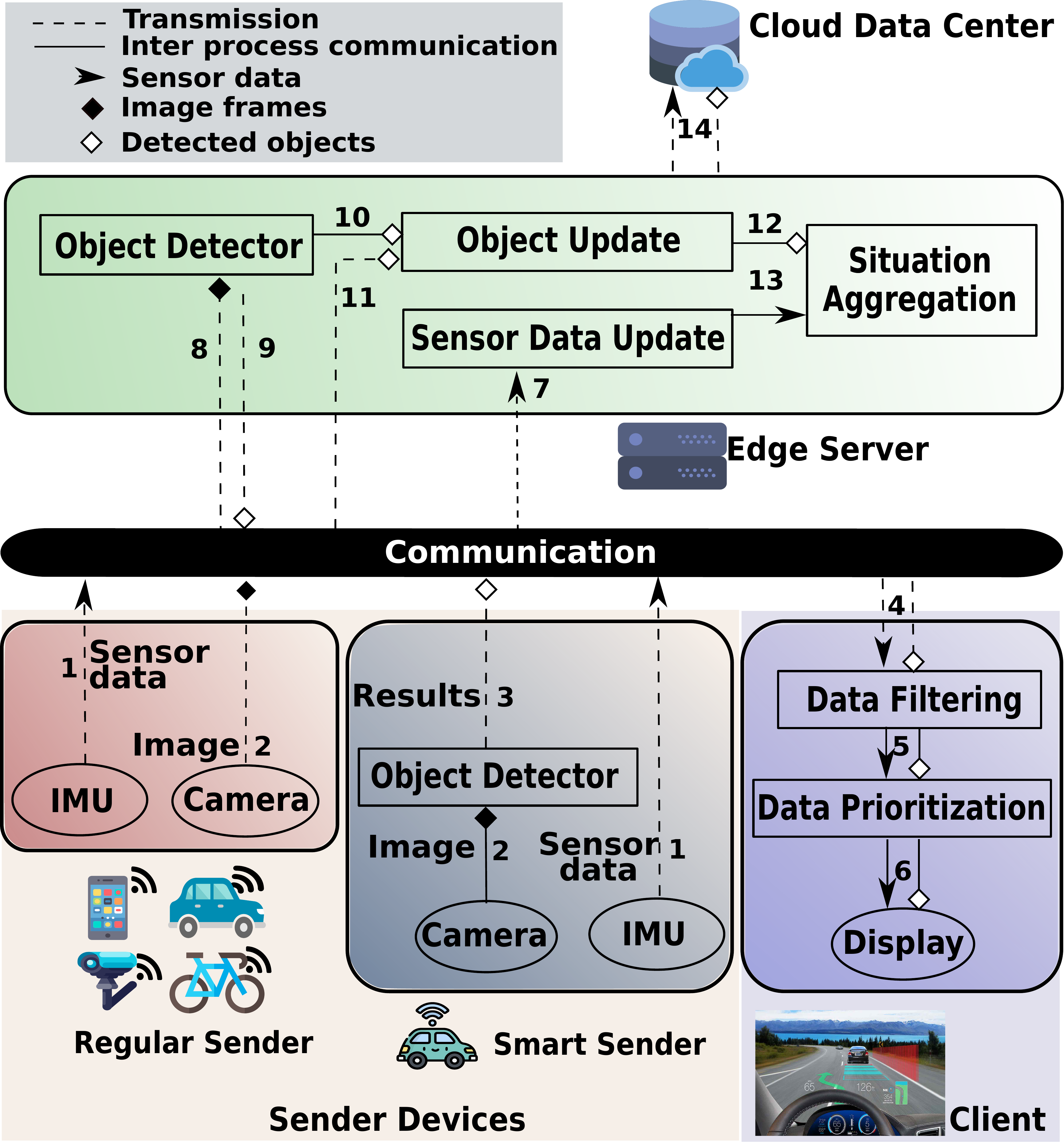}
         \caption{Edge AI offloading framework for connected vehicular perception provided by EAVVE~\cite{9163287}.}
         \label{fig:eavve}
     \end{subfigure}
     \hfill
    \begin{subfigure}[b]{0.5\textwidth}
         \centering
         \includegraphics[width=.8\columnwidth]{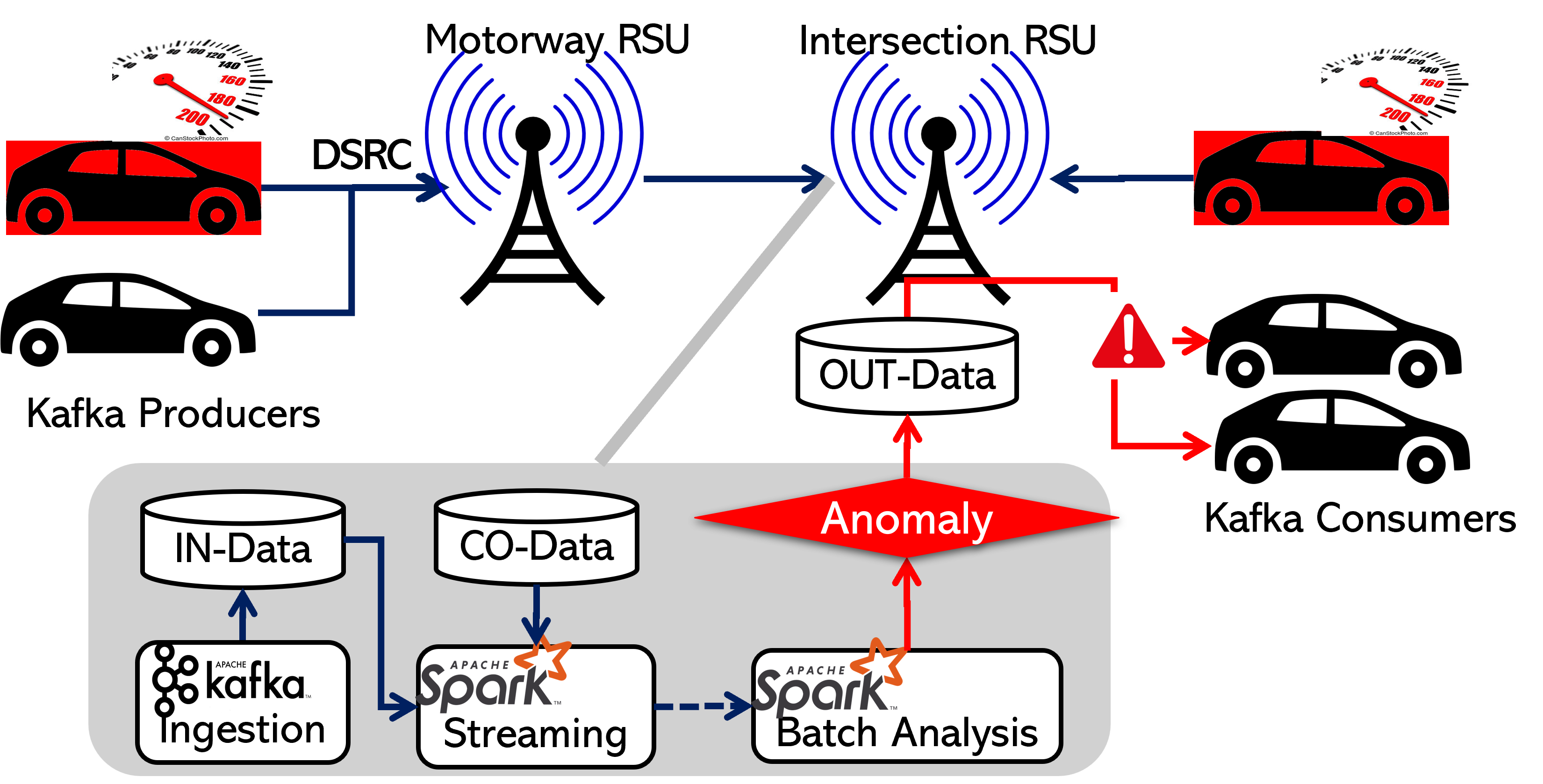}
         \caption{Unsafe driving detection in connected vehicles using RSUs by CAD3~\cite{alhilal2021cad3}.}
         \label{fig:cad3}
     \end{subfigure}
    \caption{Edge offloading to the vehicles' network edge.}
    \label{fig:edgeai}
\end{figure}

Multi-access edge computing (MEC) is anticipated to enhance the vehicular user experience by delivering standard and universal edge offloading services one hop away from connected vehicles, such as danger prediction, as 5G and 6G evolve. MEC, as proposed by the European Telecommunications Standards Institute (ETSI), is a telecommunication-vendor-centric edge cloud model in which the deployment, operation, and maintenance of edge servers are handled by an ISP operating in the area and typically co-located or one hop away from the base stations \cite{mohan2018anveshak}. It not only decreases the round-trip time (RTT) of packet delivery \cite{zhou20215g}, but it also enables near real-time orchestration for V2X interactions \cite{zhou2019enhanced,wu2021resource}. MEC is important for connected vehicular services to understand the specific local context and coordinate close cooperation between adjacent vehicles. 5G MEC servers, for instance, may manage the AR content of adjacent vehicles with a single hop of packet transfer and enable real-time cooperative AR vision. Figure \ref{fig:eavve} depicts an illustration of a MEC solution for connected vehicular vision offered by EAVVE \cite{9163287}.
Another similar solution is CAD3~\cite{alhilal2021cad3}, which integrates Apache Kafka and Spark in the RSUs to process vehicle data and detect any deviation from normal driving speed in real-time (i.e., end-to-end latency $<$50 ms) as depicted in Figure~\ref{fig:cad3}. Using AR and the location information of the vehicle, being unsafely driven, these warnings can be augmented to the vehicle's windshield. 

\section{IV-Metaverse}\label{sec:vehicle}
Besides operating the Vetaverse as a giant TS-Metaverse as described Section~\ref{sec:its}, each individual vehicle can become a small metaverse space that hosts exciting immersive services to the users inside. In this section, we present two promising services that most likely will be employed and already generate research interests, i.e., AR-RecSys and Volumetric streaming.  
\subsection{AR-RecSys}
The recommendation system (RecSys) is a vital service in the IV-Metaverse, primarily used to improve the driver's and passengers' commute experience. At present, some companies have started exploring its potential, such as BMW, Porsche, Audi, NIO, and AutoNavi map. Rich vehicle terminals and massive driving data can provide long-lasting power for the development of onboard RecSys.

Adomavicius et al.~\cite{adomavicius2005toward} have sorted out the modes of several traditional RecSys. Onboard RecSys is a new variant which combines content-based and context-aware methods, preferably augmented by knowledge-based techniques. For example, Burke et al.~\cite{burke2000knowledge} used some domain knowledge about restaurants, cuisines, and foods to recommend restaurants to users. Sun et al.~\cite{sun2009context} proposed a hierarchical context model for information definition and classification in a smart car. The recommendation of restaurants and points-of-interest~(POI) is important for the onboard RecSys. Some works~\cite{habib2016location,marques2016mobile,sojahrood2021poi,park2008restaurant,wang2021restaurant} proposed mobile recommendation systems suitable for onboard RecSys, based on features like location, time, preference, and user profiles.

\begin{figure}[t!]
    \centering
    \includegraphics[width=\columnwidth]{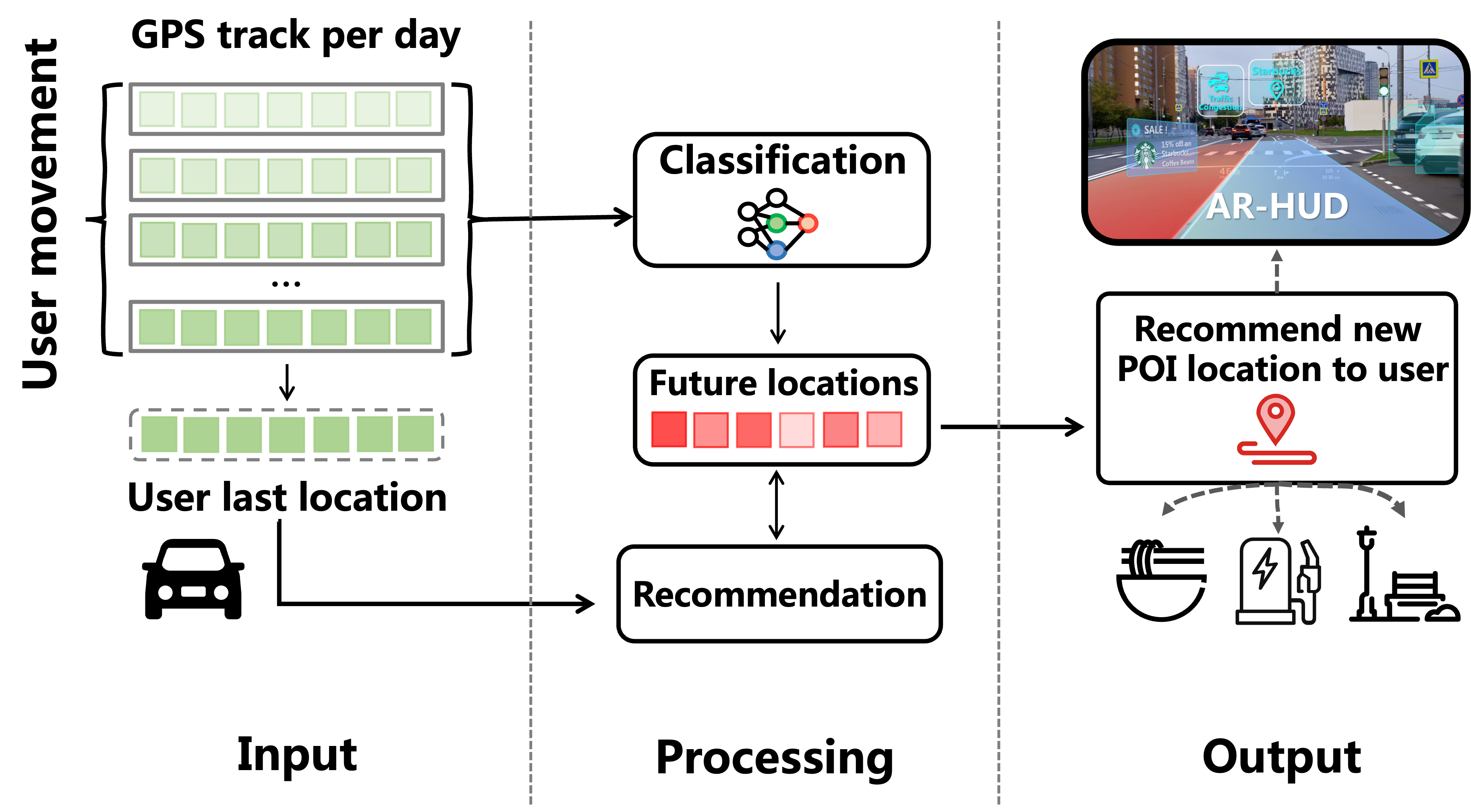}
    \caption{The general process of onboard RecSys. Three stages, (i) GPS data collection stage, data can be stored locally or uploaded to the cloud (ii) model training and calculation stage, (iii) result output stage, used to filter and display onboard recommended results.}
    \label{fig:recsys}
\end{figure}

For ITS, \cite{liu2014real,sarker2019vehicle,wu2014agile,anagnostopoulos2021predictive} proposed a variety of solutions based on location-based social networks (LBSNs) methods for route and parking recommendation. For instance, \cite{liu2014real} developed a novel route RecSys to provide self-driving tourists with real-time personalized route recommendations to reduce traffic jams and queuing time in hot spots. Sarker et al.~\cite{sarker2019vehicle} designed and implemented a multi-preference routing system VRT to recommend the optimal route with low energy consumption. Wu et al.~\cite{wu2014agile} designed an agile urban parking recommendation service to facilitate city citizens with fully efficient, real-time, and precise parking lot guiding suggestions. \cite{anagnostopoulos2021predictive} provided next destination prediction as well as a onboard RecSys based on riders’ personalized information. Figure~\ref{fig:recsys} shows the general process of the onboard RecSys in the smart city commuting scenario.

XR can greatly improve onboard RecSys with richer and personalized vehicle-human interaction. A key player in this direction is multimodal information fusion. UMPR~\cite{9152003} is a deep multimodal preferences-based recommendation method which captures the textual and visual matching of users and items for recommendation. In IV-Metaverse, the multimodal information gathered via V2V communications can provide the system with richer background and better performance. Nissan proposed Omni-Sensing technology, which uses a virtual hub to gather real-time data from the vehicle's surroundings and interior, such as road status, visibility, signage, nearby cars and pedestrians, the driver's level of alertness, and facial expressions, and body tracking for virtual avatar or Virtual Personal Assistant (VPA) interaction.

\subsection{Volumetric Streaming}
In recent years, video streaming services have become an irreplaceable part of in-vehicle infotainment services. This forces network operators and service providers to ensure a certain quality of experience (QoE) for video streaming. However, the dynamic mobility of the vehicles leads to unstable connectivity and poses variable network conditions. Therefore, adaptive video streaming is essential to deliver a high-quality experience. Many studies have been conducted to ensure satisfactory QoE in the presence of such conditions. Nebula~\cite{alhilal2022nebula} protects against visual distortion resulting from packet loss and ensures low end-to-end latency to guarantee QoE. The works \cite{eshraghi2016qoe,ghoreishi2016power,shao2015joint,jin2018qoe,li2015qoe,wang2016qoe} enhance the QoE in single-cell or multi-cell network scenarios via power control and resource allocation, while  \cite{campolo20175g,khan2021network,garcia2019performance,campolo2018towards,khan2020enhancing} satisfy the QoE using network slicing in vehicular networks.

\begin{figure}[t!]
    \centering
    \includegraphics[width=\columnwidth]{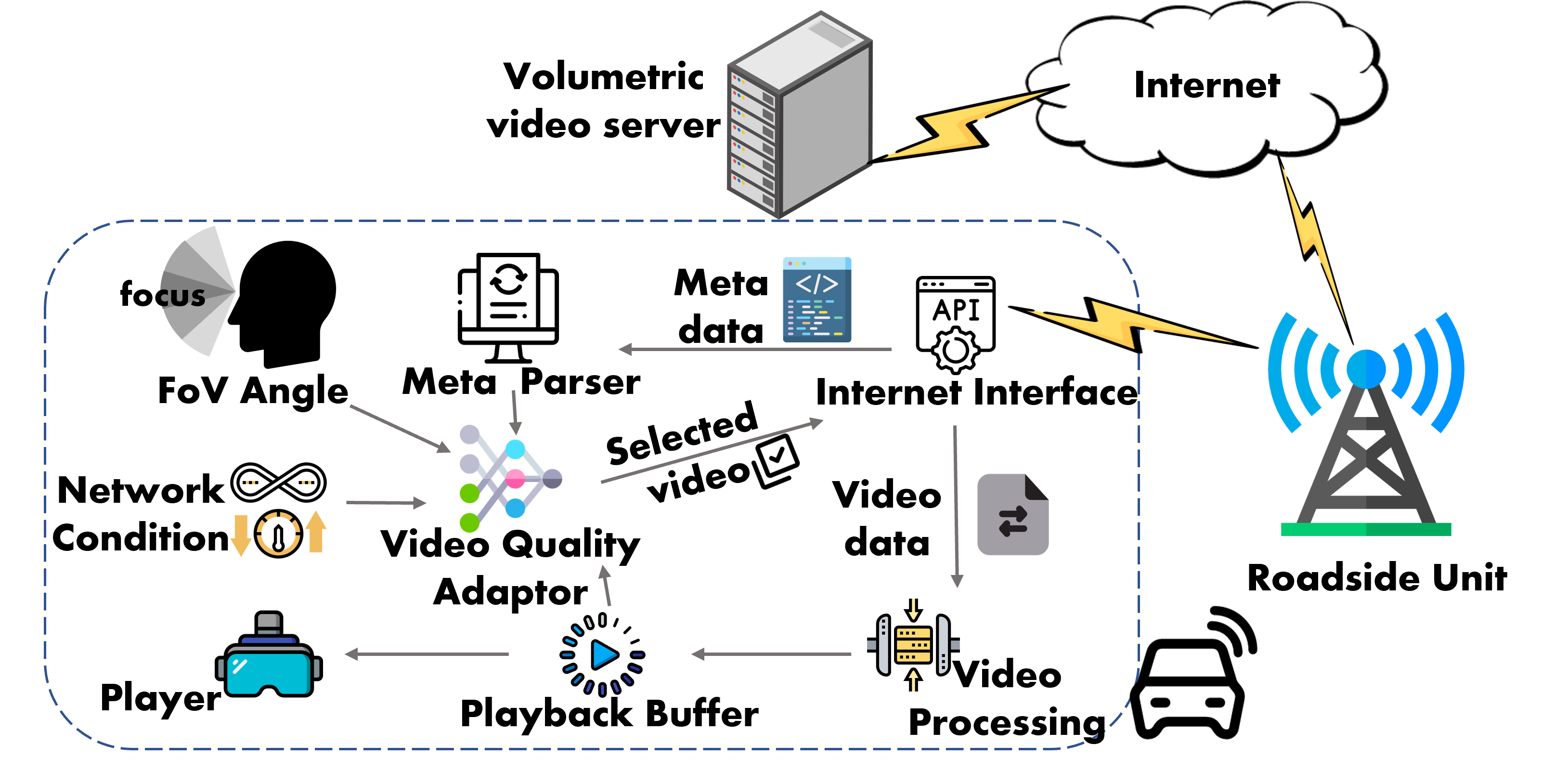}
    \caption{Vehicular adaptive volumetric video streaming framework. Considering the network bandwidth, 6DoF pose and playback buffer state to choose a proper quality chunk and receive it from the video server via the roadside unit.}
    \label{fig:streaming}
\end{figure}

Abovementioned solutions target traditional video streaming without considering the provision of an immersive experience. Volumetric streaming offers an immersive viewing experience with six degrees of freedom, i.e., 6 DoF, including the position (X, Y, Z) and the orientation (yaw, pitch, roll) of the viewer. Owing to its high encoding/decoding complexity, high video bitrate, and low latency requirement, the transmission of volumetric video remains a challenging problem. Van der Hooft et al.~\cite{van2019towards} propose PCC-DASH, a standards-compliant means for HTTP adaptive streaming of scenes, comprising multiple, dynamic point cloud objects. They present several rate adaptation heuristics using users' position and focus, the available bandwidth, and the client’s buffer status to decide upon the most appropriate quality representation of each object. ViVo~\cite{han2020vivo} performs the first comprehensive study of mobile volumetric video streaming. To reduce bandwidth consumption, ViVo determines the video content to fetch based on how, what, and where the viewer is perceiving it. This solution can save on average 40\% of data usage with virtually no drop in visual quality according to their evaluations. PCC-DASH and ViVo adapt the video bitrate using heuristic algorithms. FRAS~\cite{gao2022fras} proposes the first federated reinforcement learning framework for adaptive point cloud video streaming based on a variety of features, including the network conditions, the viewer’s focus and position, playback buffer state, and the device computing capability etc. FRAS enables joint training using multiple participants and preserves privacy while enhancing the training performance. Figure~\ref{fig:streaming} illustrates the framework of adaptive volumetric video streaming in vehicular networks.

In addition to the traditional adaptive bitrate (ABR) algorithms, there are numerous techniques to optimize vehicular volumetric video streaming. For instance, Anlan Zhang, et al.~\cite{zhang2022yuzu} propose to leverage 3D super-resolution (SR) to drastically increase the visual quality of volumetric video streaming. Lee et al~\cite{lee2020groot} enhance the viewer’s QoE by optimizing the transmission formats and codec methods of the volumetric video. Yakun Huang, et al.~\cite{huang2021aitransfer} propose AITransfer, an AI-powered bandwidth-aware and adaptive transmission technique driven by extracting and transferring key point cloud features to reduce bandwidth consumption and alleviate computational pressure.


\subsection{Virtual Companion}
An emotional avatar is essentially a digital representation through VR methods, which can adaptively adjust itself to the appropriate appearance according to the user's emotional situation. In more detail, an emotion avatar expresses corresponding emotions based on the user's emotion recognition to alleviate the user's negative emotions. Emotion avatars have already been applied in some areas. FaceSay~\cite{hopkins2011avatar} improves children's social skills with interactive, realistic emotion avatar assistants. ~\cite{8172404} proposed a robotic system as an emotional avatar which detects the user's emotions and enacts them to extend users' methods of communication with motor disabilities. 

\begin{figure}[t!]
    \centering
    \includegraphics[width=\columnwidth]{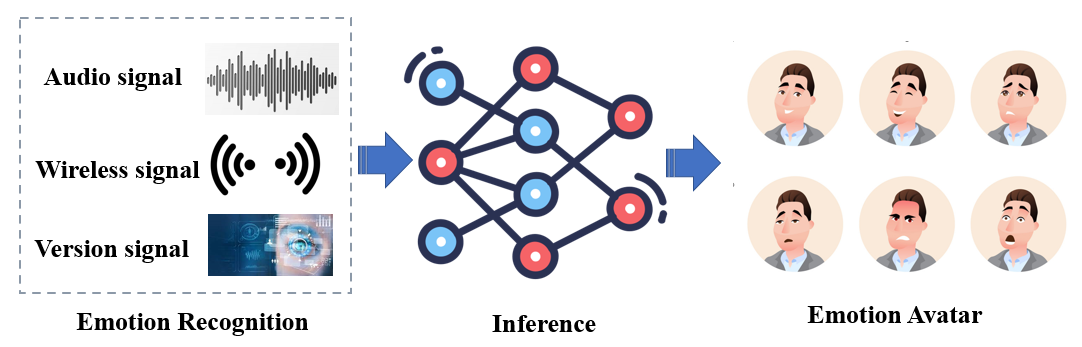}
    \caption{Emotion avatar framework. By recognizing the emotional state of the user, the emotion avatar will show appropriate performance after comprehensive judgment to  soothe the user's emotions, avoiding accidents caused by bad emotions.}
    \label{fig:emotionavatar}
\end{figure}

Figure~\ref{fig:emotionavatar} shows how the emotional avatar framework works in IV-Metaverse. Introducing emotional avatars can potentially improve the user's emotional experience and driving safety.

\subsubsection{Spatial Audio for ADAS}
Back in 2002, Holland et al.~\cite{hollandAudioGPSSpatialAudio2002} proposed AudioGPS, the first spatial audio-based global positioning system (GPS) with a user interface. This was intended to allow a mobile computer to perform location tasks when the user's attention is otherwise engaged. AudioGPS accepts the audio signal and converts it into a binaural signal that the user listens to through headphones. The system then provides an audio representation of the direction and distance to the destination for navigation. In 2016, Albrecht et al.~\cite{albrecht2016guided} creatively integrated the enjoyment of music with navigation services. They use spatial audio for navigation by way of route guides and beacon guides. Drivers can drive to their destination according to the directions indicated by the music. In 2018, Heller et al.~\cite{heller2018navigatone} proposed NavigaTone, a system that utilizes multichannel recording and provides directional navigation by moving individual tracks in auditory space. The driver can position the sound source as if using stereo panning, while the listening experience is closer to that of regular music listening. 

In 2019, Kari et al.~\cite{kari2021soundsride} applied the idea of procedural game music to car drivers using a system called Soundsride. It is an onboard audio AR system that synchronizes music with the environment in real time. In the system, high-contrast events in music are aligned with high-contrast events in the environment as the vehicle enters or exits a sound affordance. It also uses recursive filtering algorithms to keep the temporal availability distance up to date and change the audio signal in real time. The results prove that Soundsride can create a great music experience and positively impact subjective driving safety. In 2021, Dupre et al.~\cite{dupreSpatialSoundDesign2021} achieved the synthesis of spatial audio and the integration of ambient sounds in the vehicle cockpit by adjusting timbre and spatial parameters. They did this by combining simultaneous 3D sound field measurements with 3D video recording in the seat view, allowing the driver to actually control the sound source from inside the vehicle and to be spatially aware of the overall vehicle design soundscape. This approach creates a suitable and pleasant soundscape inside the vehicle to ensure the user's comfort and safety.

\section{Manufacturing}
IV-Metaverse and TS-Metaverse are two kinds of ``runtime'' Vetaverse. Industries have recognised the enormous potential of using XR techniques in improving the performance and efficiency of manufacturing flows. Promising deployments include digital twin factories, XR for design, XR for maintenance and diagnosis, XR for show and training, and so on. This section lays out these directions and describes current developments.

\subsection{Digital Twin Factory}

The new generation of information and intelligent technology is taking up an increasingly important position in the manufacturing industry, and various countries around the world have put forward various strategic plans to promote the development of digital transformation of  manufacturing industries. Digital twin factory plays an important role in this trend~\cite{tao2018digitalTwin, grieves2015digitalTwin}

The digital twin factory refers to that after establishing virtual models for the equipment in the production line and associating the physical entities with the virtual models through sensors, IoT and other technologies, useful data can be obtained from multiple levels and angles through relevant software devices and networks, and the factory can be continuously and comprehensively monitored and managed in terms of status. In turn, based on the calculation and prediction of the data, the sharing and scheduling of resource elements through the digital twin platform will promote the advancement from local business optimization such as digital design and intelligent production to global resource collaboration and optimization such as networked collaboration and shared manufacturing~\cite{china2022factory}.

\subsection{XR for Design}

Vehicle design is a long process requiring iterative modifications and reviews with back-and-forth cycles. It is an expensive and time-consuming process that demands lots of manual effort to propose ideas, implement mock-ups, discuss and review, revise, and so on. Using VR to speed up the implementation and review processes has been investigated for the past 20 years. Virtual prototypes can visualize and identify problems at a very early stage to reduce the costs and life cycle of design. In 1997, Lehner and De Fanti~\cite{lehner1997distributed} built a VR software with networked communication capacity to test collaborative virtual prototyping among remotely located engineers. Coburn, Freeman and Salmon~\cite{Coburn2017review} concluded that there were significant opportunities for using VR during design tasks to improve results and reduce development time. Recently, in 2021, the development, experimental validation, and use of a digital twin for an automotive traction drive system was illustrated by Liu et al~\cite{liu2021digital}

\subsection{XR for Maintenance and Diagnosis}

XR can also provide helpful assistance for equipment maintenance and diagnosis in the manufacturing sector. As a result of the great distances between businesses and equipment suppliers, this previously required expert consultation and diagnostics on-site, which came at a cost to businesses in addition to the high cost of travel. XR technology and the digital twin platform enable remote equipment condition viewing and diagnosis, monitoring and aggregation of various data information, and assisted diagnosis of equipment status. In today's globalized trade and manufacturing sectors, this can effectively lower the cost of labor and materials, and equipment maintenance, lower the loss from equipment failure, and thus effectively improve the overall efficiency of the factory.

On the other hand, XR can offer three-dimensional visual guidance during routine maintenance, enabling staff to prevent missing parts that need to be checked due to time constraints. The main benefits of using computer-based systems for technician training or support are that computers don't forget things and that they can make information easier for people to understand. These features can aid in reducing errors brought on by following procedures incorrectly, interpreting information incorrectly, or lacking sufficient training~\cite{Francesca2011Augmented}. 

\subsection{XR for Auto Show}

The use of cloud-based resources and vehicle 3D modeling can realize high-definition, real-time, interactive, and immersive online virtual car shows, such as combining vehicle data with vehicle entities in AR, which helps customers have a more comprehensive and thorough understanding of vehicles. This is made possible by the development of extended reality rendering capabilities and the digital transformation of the automotive industry. Simultaneously, online car shows overcome traditional time and space constraints, allowing customers to view and experience different model configurations at their leisure, reducing customers' time overhead, significantly improving their show experience, and increasing their willingness to buy.

\subsection{XR for Training}

XR is also handy for personnel training. Traditional training methods necessitate extensive specialized training and on-site guidance from skilled personnel in order to become acquainted with operations, but such methods have the following drawbacks: 1. The overhead on human and equipment resources is high; 2. Paper teaching is not intuitive enough, and training efficiency is low; and 3. Direct contact with the production environment poses potential safety risks. 4. One-to-many detailed guidance is difficult to achieve, and there may be omissions in teaching.

In response to these problems, XR technology can improve the training process from multiple perspectives~\cite{DORNELLES2022107804,10.1007/978-3-319-64465-3_43,doi:10.1080/0951192X.2015.1067918,Mao2022}. For instance, using MR equipment to direct trainees to perform correct and thorough operations and using virtual reality equipment to simulate real scenes are both effective ways to increase training effectiveness and safety. It can also make use of the network to exchange better instructional materials, prevent omissions during training, and impart knowledge in a way that helps students form virtuous operating habits. As an example, Rivera et al.~\cite{Rivera2020training} designed an augmented reality training system for hybrid vehicles, and the experimental findings show a high rate of acceptability towards the augmented reality training system since users can manipulate the THS's primary components, cutting down on the amount of time spent practicing, and enhancing student learning through secure working environments.

\section{Challenges and Open Research Directions}\label{sec:discussion}
The development and deployment of Vetaverse on large scale face many challenges regarding resource and operation efficiency, timely data delivery, privacy and security, safety and well-being, etc. Here we discuss three challenges in detail.

\subsection{High Capacity Communication}
Vetaverse requires high bandwidth and low latency communication to deliver audio-video services in real-time for an immersive experience (e.g., $360^0$ VR video delivery)~\cite{xiao2022transcoding}. It is anticipated that current communication systems (LTE and 5G) will fail to satisfy these requirements. While mobile edge computing (MEC) is a crucial component for the computation offloading of users' heavy tasks to satisfy low-latency demands, the bandwidth offered by the 4G and 5G networks could be incapable of meeting Vetaverse demand. The sixth generation (6G) networks are envisioned to enable Vetaverse by providing the end users with high-capacity communications to the MEC servers~\cite{chang20226g}. Future work can study the utilization of 6G networks to cover a variety of deployment use cases and applications of Vetaverse. This would involve investigating emerging 6G technologies such as visible light communication (VLC) and Terahertz (THz) wireless communication, intelligent reflecting surfaces (IRS), and non-orthogonal multiple access (NOMA). This would also involve studying deployment use cases to cope with data traffic congestion.

\subsection{Safety}
A user's awareness and understanding can be improved by the rich information provided in XR. However, if the interface and information delivery are poorly designed, it may potentially draw the driver's attention and present a threat. Therefore, it is essential to carefully design the onboard XR interfaces that run the danger of causing distractions and carry out exhaustive user research and road tests. For instance, when displaying too much or in the incorrect places, ARHUD, which overlays augmented reality information on the real world, can easily obstruct a driver's perspective. Data filtering~\cite{aicp} and user-friendly display are therefore essential. When users look at the emotional avatar on the central control panel, which is where we envision it being rendered to intimately engage with the drivers and passengers, it may lead to the driver's head movements, which severely impact the driving safety. Therefore, spatial audio~\cite{kari2021soundsride} facilitated techniques are important to help the driver avoid frequent head movements.

\subsection{Human-centered Trustworthiness}
Rapidly developing technologies are getting more and more potent, raising questions about their reliability from a human perspective. Both XR and AI are expanding quickly and finding more use in people's daily lives. Therefore, it is essential to guarantee that both their basic operating processes and the services they provide are reliable. Prior to now, the reliability and robustness of systems and algorithms was more of a priority in relation to the primary and original aim of the design. As suggested by numerous AI ethics guidelines around the world~\cite{hleg2019ethics,usdod}, the emphasis is now turning to human-centered considerations, such as fairness, privacy, explainability, and well-being among others. Both Vetaverse and Metaverse are struggling with difficulties of trustworthiness issues, because users are more likely to become confused in the immersive blended virtual-reality world. Therefore, the primary requirements and challenges are to ensure that the immersive services are benign, reliable, and trustworthy.

\section{Conclusion}\label{sec:conclusion}
In this paper, we define the anticipated Vetaverse, which is the meeting point of the Metaverse with the world of cars and transportation. We suggest the taxonomy and thoroughly outline the major enablers. We divide the Vetaverse into two subcategories—IV-Metaverse and TS-Metaverse—and outline a number of potential immersive, futuristic services they might offer. We also go through some of the most important unsolved problems for interested scholars.

\bibliographystyle{plain}
\bibliography{reference}

\end{document}